\documentclass[%
 reprint,
superscriptaddress,
 amsmath,amssymb,
 aps,
]{revtex4-2}

\usepackage{graphicx}% Include figure files
\usepackage{dcolumn}% Align table columns on decimal point
\usepackage{bm}% bold math
\newcommand{\sigvec}{\vec{\sigma}}
\usepackage{braket}
\newcommand{\superket}[1]{\ket{#1}\rangle}
\newcommand{\superbra}[1]{\langle\bra{#1}}
\usepackage{comment}
\usepackage{hyperref}

\begin{document}

\preprint{APS/123-QED}

\title{
An automated geometric space curve approach for designing \\ dynamically corrected gates}% 

\author{Evangelos Piliouras}
 \affiliation{Department of Physics, Virginia Tech, Blacksburg, VA 24061, USA}
\affiliation{Virginia Tech Center for Quantum Information Science and Engineering, Blacksburg, VA 24061, USA}
\author{Dennis Lucarelli}
\affiliation{Error Corp., College Park, MD 20740, USA}
\author{Edwin Barnes}
 \affiliation{Department of Physics, Virginia Tech, Blacksburg, VA 24061, USA}
\affiliation{Virginia Tech Center for Quantum Information Science and Engineering, Blacksburg, VA 24061, USA}

\date{\today}% It is always \today, today,

\begin{abstract}
The noisy nature of quantum hardware necessitates the implementation of high-fidelity quantum gates in a noise-insensitive manner. While there exist many powerful methods for designing dynamically corrected gates, they typically involve an exploration across a large-dimensional landscape filled with solutions that are only locally optimal, making it challenging to find globally optimal ones. Moreover, these methods often use a single cost function to try to accomplish the two disparate goals of achieving a target gate and suppressing noise, and this can lead to unnecessary tradeoffs between the two and, consequently, lower fidelities. Here, we present a method for designing dynamically corrected gates called Bézier Ansatz for Robust Quantum (BARQ) control to address these challenges. Rather than numerically optimizing the controls directly, BARQ instead makes use of the Space Curve Quantum Control formalism in which the quantum evolution is mapped to a geometric space curve. In the formulation used by BARQ, the boundary conditions of the space curve determine the target gate, while its shape determines the noise robustness of the corresponding gate. This allows the target gate to be fixed upfront, so that numerical optimization is only needed to achieve noise-robustness, and this is performed efficiently using a control-point parameterization of the space curve. In this way, BARQ eliminates the gate-fixing and noise-robustness tradeoff while also providing a global perspective into the control landscape, and allows for ample freedom to design experimentally friendly and robust control pulses. The pulse design is facilitated through the developed software package \texttt{qurveros}.
\end{abstract}
\maketitle
\section{Introduction}
Fault-tolerant scalable quantum computation demands the ability to implement quantum gates at error rates that are well below a certain error threshold \cite{
AHARONOVFaultTolerantQuantumComputationConstant2008,
KNILLResilientQuantumComputationError1998,
PRESKILLReliableQuantumComputers1998,
SANDERSBoundingQuantumGateError2015}. Achieving this requires not only improvements in quantum hardware, but also in the control waveforms that are used to execute gates. Infinitely many control waveforms can generate the same target gate, but only some choices will do so in a way that is robust to noise and other sources of error in the hardware. Finding these preferred choices is the goal of robust quantum optimal control methodologies.

Robust quantum control techniques have a long history dating back to the early 1950s and have achieved tremendous success in various use cases that include dynamical decoupling (DD) \cite{
BIERCUKDynamicalDecouplingSequenceConstruction2011,
BIERCUKOptimizedDynamicalDecouplingModel2009,
CYWINSKIHowEnhanceDephasingTime2008,
KHODJASTEHFaultTolerantQuantumDynamicalDecoupling2005,
NGCombiningDynamicalDecouplingFaulttolerant2011,
TRIPATHIQuditDynamicalDecouplingSuperconducting2024,
UHRIGExactResultsDynamicalDecoupling2008,
VIOLADynamicalDecouplingOpenQuantum1999} and dynamically corrected gates (DCGs) \cite{KHODJASTEHAutomatedSynthesisDynamicallyCorrected2012,
KHODJASTEHArbitrarilyAccurateDynamicalControl2010,
KHODJASTEHDynamicallyErrorCorrectedGatesUniversal2009,
KHODJASTEHDynamicalQuantumErrorCorrection2009,BROWNArbitrarilyAccurateCompositePulse2004,
KABYTAYEVRobustnessCompositePulsesTimedependent2014}. Meanwhile, optimal quantum control theory (OCT) strategies \cite{ANSELIntroductionTheoreticalExperimentalAspects2024,
BOSCAINIntroductionPontryaginMaximumPrinciple2021,
CANEVAOptimalControlQuantumSpeed2009,
CANEVAChoppedRandombasisQuantumOptimization2011,
DALESSANDROIntroductionQuantumControlDynamics2021,
DEFFNERQuantumSpeedLimitsHeisenbergs2017,
DORIAOptimalControlTechniqueManyBody2011,
GLASERTrainingSchrodingersCatQuantum2015,
JAGEROptimalQuantumControlBoseEinstein2014,
KHANEJAOptimalControlCoupledSpin2005,
KOCHQuantumOptimalControlQuantum2022,
LUCARELLIQuantumOptimalControlGradient2018,
MACHNESTunableFlexibleEfficientOptimization2018,
MULLEROneDecadeQuantumOptimal2022,
POGGIQuantumSpeedLimitOptimal2013,
REICHMonotonicallyConvergentOptimizationQuantum2012,
SKLARZLoadingBoseEinsteinCondensateOptical2002,
SORENSENQuantumOptimalControlChopped2018}
have been effective in solving the problem of high-fidelity state transfer, where each method is unique in terms of the parameterization of the control problem and the search strategy towards the optimal solution. The search can be carried out using analytical gradients of cost functions \cite{MACHNESTunableFlexibleEfficientOptimization2018}, with traditional approaches requiring the discretization of the pulse into constant segments \cite{KHANEJAOptimalControlCoupledSpin2005}, or optimization can be performed in a gradient-free manner \cite{CANEVAChoppedRandombasisQuantumOptimization2011}. The choice of the expansion basis and their suitability to a particular goal remains an open problem
\cite{PAGANORoleBasesQuantumOptimal2024,CAROLANRobustnessControlledHamiltonianApproaches2023}. 

In practice, the design of optimal robust quantum control protocols is achieved by defining cost functions that enforce a particular condition upon minimization. When a multitude of constraints is required, the optimization will naturally prefer to minimize the quantity with the highest weight. Such a situation is exacerbated when experimental constraints must be taken into account \cite{ANSELIntroductionTheoreticalExperimentalAspects2024}. A characteristic example is the tradeoff between reaching a target state in minimal time (in the ideal case) and whether this is performed in a noise-insensitive manner (robustness). Although Ref.~\cite{POGGIUniversallyRobustQuantumControl2024} showed that this tradeoff can in principle be avoided, in practice it still arises frequently due to a combination of how the problem is formulated mathematically and which optimization strategy is chosen. Despite the tremendous progress in both domains, the computationally efficient, systematic, and tradeoff-free design of high-fidelity, noise-robust, and experimentally friendly DCGs still remains an outstanding challenge.

In an effort to address this challenge, several works developed robust quantum control protocols \cite{GLASERTrainingSchrodingersCatQuantum2015,
KOCHQuantumOptimalControlQuantum2022,
WEIDNERRobustQuantumControlClosed2024}
that utilize experimentally feasible pulses. The proposed methods vary in the degree to which analytical information is injected into the design procedure and in the definition of optimality considered in each case. Methods that incorporate more analytical information can more easily find globally robust solutions by narrowing the search space~\cite{BARNESRobustQuantumControlUsing2015,
ZENGFastestPulsesThatImplement2018,
ZENGGeneralSolutionInhomogeneousDephasing2018,
COLMENARConditionsEquivalentNoiseSensitivity2022,
DRIDIRobustControlNotGate2020,
DRIDIOptimalRobustQuantumControl2020,
GUNGORDUAnalyticallyParametrizedSolutionsRobust2019,
HANSONConstructingNoiseRobustQuantumGates2024,
HARUTYUNYANDigitalOptimalRobustControl2023,
MEINERSENQuantumGeometricProtocolsFast2024,
RIMBACH-RUSSSimpleFrameworkSystematicHighfidelity2023,
WANGRobustQuantumGatesSinglettriplet2014,
VANDAMMEApplicationSmalltipangleApproximationToggling2021,
KHROMETSQuantumOptimalControlRobust2024}, but they can also prove more challenging to work with when multiple constraints need to be imposed. On the other hand, more numerical quantum optimal control methods can more easily produce control pulses that satisfy a large set of constraints \cite{COLMENARReverseEngineeringOnequbitFilter2022,
KOSUTRobustQuantumControlAnalysis2022,
LEAnalyticFilterFunctionDerivativesQuantum2022,
LEUNGSpeedupQuantumOptimalControl2017,
LUCARELLIQuantumOptimalControlGradient2018,
MOTZOIOptimalControlMethodsRapidly2011,
MULLEROneDecadeQuantumOptimal2022,
ODAOptimallyBandLimitedNoiseFiltering2023,
POGGIUniversallyRobustQuantumControl2024,
RIBEIROSystematicMagnusBasedApproachSuppressing2017,
ROSSIGNOLOQuOCSQuantumOptimalControl2023,
SONGOptimizingQuantumControlPulses2022,
TESKEQoptExperimentOrientedSoftwarePackage2022,
YIRobustQuantumGatesCorrelated2024,
YANGQuantumControlTimedependentNoise2024}, and they can tailor the resulting pulses to the specific system to a greater degree~\cite{BALLSoftwareToolsQuantumControl2021,
CARVALHOErrorRobustQuantumLogicOptimization2021,
GENOISQuantumOptimalControlSuperconducting2024,
GUNGORDURobustQuantumGatesUsing2022,
KOSUTRobustControlQuantumGates2013,
NIUUniversalQuantumControlDeep2019,
PROPSONRobustQuantumOptimalControl2022,
SIVAKModelFreeQuantumControlReinforcement2022}, but they can also introduce unwanted tradeoffs between key quantities of the control objective, which can in turn limit fidelities or increase gate times.

In this paper, we propose the Bézier Ansatz for Robust Quantum (BARQ) control method. Our approach makes use of the Space Curve Quantum Control (SCQC) formalism~\cite{BARNESDynamicallyCorrectedGatesGeometric2022,
BUTERAKOSGeometricalFormalismDynamicallyCorrected2021,
DONGDoublyGeometricQuantumControl2021,
LIDesigningArbitrarySingleaxisRotations2021,
NELSONDesigningDynamicallyCorrectedGates2023,
ZENGGeometricFormalismConstructingArbitrary2019,
ZHUANGNoiseresistantLandauZenerSweepsGeometrical2022}, which recasts the problem of designing DCGs to the equivalent problem of constructing space curves with certain properties, allowing us to sidestep the integration of the time-dependent Schr\"{o}dinger equation. We obtain suitable space curves by expressing them as Bézier curves and optimizing over control points until the space curve shape satisfies certain constraints. Control pulses that generate the target DCGs can then be extracted from the derivatives of the space curves. In a manner analogous to the approach taken in Ref.~\cite{LUCARELLIQuantumOptimalControlGradient2018}, where the optimization is restricted to the space of band-limited control signals, optimizing in the space of closed B{\'e}zier curves constrains the controls to be noise-suppressing \emph{a priori}. Furthermore, we reformulate SCQC in the adjoint representation, which allows us to decouple the gate-fixing and noise-robustness problems, such that the target gate is determined solely by the boundary conditions of the space curve, while noise-robustness is encoded purely in its shape. This formulation prevents any tradeoff between gate-fixing and noise-robustness from arising, and it guarantees unit gate fidelity in the absence of noise, so that only the noise-robustness property requires optimization. Our method can be regarded as a continuous interpolation of quantum optimal control methods that discretize the quantum evolution but nevertheless operate with analytical continuous gradients. We provide two examples of BARQ-optimized gates and assess their performance in two noise scenarios: simultaneous quasi-static dephasing and multiplicative driving field errors, and low-frequency time-dependent dephasing errors. Although standard formulations of SCQC can potentially generate non-smooth pulses with fast rise times, with BARQ, the experimental feasibility of the resulting pulses arises naturally from the use of Bézier curves and control points.   

The paper is structured as follows. In Sec. \ref{robust_control_intro}, we describe the goals of the robust quantum control problem and provide their equivalent formulation in the SCQC formalism. In Sec. \ref{barq_prelimis_method}, we introduce the BARQ method, providing new theoretical extensions in SCQC and deriving the conditions which the control points of the Bézier curves must satisfy. An overview of the software package \texttt{qurveros} is also provided. In Sec. \ref{results}, we design two quantum operations using BARQ and assess their performance on various types of noise. In Sec. \ref{conclusion}, we provide some conclusions.
\section{Robust Quantum Control as a space Curve design problem}
\label{robust_control_intro}
The task of designing DCGs amounts to targeting two goals simultaneously: (i) Ensuring that the target gate operation is achieved in the absence of noise (``gate-fixing"), and (ii) making sure the gate is implemented in a way that is insensitive to noise or other sources of error (``noise-robustness"). In this section, we recast this problem in terms of the SCQC formalism, which maps robust quantum trajectories onto space curves with particular geometric properties.

\subsection{Robust Quantum Control: \\Two goals, one control}

We first discuss the task of gate-fixing. We denote the ideal, noise-free Hamiltonian by $H_0$ and the evolution operator it generates by $U_0$. The latter must solve the time-dependent Schr\"{o}dinger equation, and we further require that it equals a target gate $U_g$ at a final time $T_g$:
\begin{align}
    i\dot U_0 = H_0 U_0,\qquad U_0(T_g) = U_g.
    \label{gate_fixing_goal}
\end{align}
There are generally an infinite number of ways to choose the control fields in $H_0$ such that Eq.~\eqref{gate_fixing_goal} is satisfied, and there are many methods to find explicit examples. However, only some of these control fields will also suppress noise, as we discuss next.

We will describe the noise-robustness task by first introducing an arbitrary noise operator $H_\text{n}$ into the Hamiltonian, $H = H_0 +H_\text{n}$, which gives rise to a noisy propagator $U$. We can isolate the effect of this noise term on the evolution by moving to the interaction picture or toggling frame
\cite{GREENArbitraryQuantumControlQubits2013,KHODJASTEHDynamicalQuantumErrorCorrection2009}. This is done using the decomposition $U=U_0U_I$ and then solving the effective Schr\"{o}dinger equation for $U_I$:
\begin{align}
    i\dot U_I = (U_0^\dagger H_\text{n} U_0)U_I.
\end{align}
In the absence of the noise term $H_\mathrm{n}$, we have $U_I(t)=I$, while any amount of noise will cause $U_I$ to deviate away from the identity.
A common example is when a qubit's frequency is not precisely known or exhibits fluctuations as a result of the qubit’s interaction with its environment; this can be described by a noise term of the form $H_\text{n} = \delta_z(t)\sigma_z$, where $\delta_z(t)$ is an unknown and possibly stochastic function. The leading-order effect of $H_\mathrm{n}$ is captured by the first-order term of the Magnus expansion \cite{BLANESMagnusExpansionItsApplications2009}:
\begin{align}
    U_I \approx e^{-iH_\text{eff}T_g},\qquad H_\text{eff} = \frac{1}{T_g}\int_0^{T_g} dt\, U_0^\dagger H_\text{n} U_0.
\end{align}
which yields the (approximate) toggling frame error dynamics in the context of the Average Hamiltonian Theory \cite{BRINKMANNIntroductionAverageHamiltonianTheory2016}.
From this perspective, noise-robustness can be achieved by minimizing the average (unitless) Hamiltonian quantity:
\begin{align}
  \left|\left| \frac{1}{T_g} \int_0^{T_g} dt\, U_0^\dagger (T_g H_\text{n})U_0 \right|\right|_2^2,
  \label{pulse_robustness_goal}
\end{align}
where the operator norm of an operator $A$ is defined as $\|A\|_2^2 = \text{tr}(A^\dagger A)$.
From this expression, it is apparent that a necessary (but not sufficient) condition for noise suppression is $[U_0, H_{\text{n}}] \neq 0$.

\subsection{Brief overview of\\ Space Curve Quantum Control}

There exist a variety of methods to find control Hamiltonians $H_0(t)$ that satisfy both gate-fixing, Eq.~\eqref{gate_fixing_goal}, and noise-robustness, Eq.~\eqref{pulse_robustness_goal}. The central idea is to optimize the control pulses in $H_0$ such that both goals are achieved within a desired accuracy. In this section, we will recast these goals into the language of differential geometry to facilitate the process of finding optimal pulse shapes. 

The description of quantum evolution in terms of geometric quantities dates back nearly seventy years. Feynman et al.~\cite{FEYNMANGeometricalRepresentationSchrodingerEquation1957} associated the complex coefficients of a two-level system wavefunction with the trajectory of a time-evolving Bloch vector~\cite{BALAKRISHNANClassicalAnaloguesSchrodingerHeisenberg2004,
DANDOLOFFTwolevelSystemsSpaceCurve1992,
DANDOLOFFParallelTransportSpaceCurve1989}. There is not a unique way to map quantum dynamics to geometrical paths; one should choose a mapping that best elucidates the aspects of the quantum evolution one is interested in~\cite{CARMELGeometricalApproachTwolevelHamiltonians2000,CAFAROCurvatureQuantumEvolutionsQubits2024,RANGELOVStimulatedRamanAdiabaticPassage2009}. 

Space Curve Quantum Control (SCQC)~\cite{BARNESDynamicallyCorrectedGatesGeometric2022} is a framework for designing DCGs that maps quantum evolution to geometric space curves such that noise-robustness has a natural geometric interpretation. In particular, SCQC defines the space curve such that Eq.~\eqref{pulse_robustness_goal} vanishes precisely when the space curve closes on itself. For example, when a single qubit is subject to quasistatic dephasing noise with $H_\mathrm{n}\sim\sigma_z$, achieving robustness to first order is equivalent to requiring
\begin{align}
    \int_0^{T_g}dt \, U_0^\dagger \sigma_z U_0 = 0.
\end{align}
In SCQC, the space curve $\vec{r}(t)$ is thus defined by
\begin{align}
    \int_0^{t} dt' \, U_0^\dagger \sigma_z U_0 = \vec r (t) \cdot \sigvec,
    \label{scqc_main_eq}
\end{align}
where $\sigvec = \begin{bmatrix}
    \sigma_x & \sigma_y & \sigma_z
\end{bmatrix}^T$ is a vector of Pauli matrices, and the $\cdot$ operation is the inner product between two vectors. With this definition of $\vec{r}(t)$, robustness against dephasing becomes the condition that $\vec r(T_g) = \vec r(0)$, which means that the space curve is closed~\footnote{In this paper, we use the term closed curve to refer to a curve whose initial and final points coincide. Note that the term is sometimes given another meaning in the differential geometry literature.}. Any closed curve we construct will correspond to an evolution operator that is robust to dephasing to first order. More generally, if $H_\mathrm{n}$ contains multiple independent noise sources, then we choose one of these terms to define the space curve, while the cancellation or suppression of the remaining noise terms translates to additional constraints on the space curve beyond its closure~\cite{NELSONDesigningDynamicallyCorrectedGates2023}.

Once we have constructed a space curve that satisfies the robustness condition, we can compute its derivatives to find the control Hamiltonian $H_0(t)$ that generates this evolution. While SCQC can be applied to systems with any finite number of Hilbert space dimensions~\cite{BUTERAKOSGeometricalFormalismDynamicallyCorrected2021}, in this work we focus on a general single-qubit Hamiltonian of the form
\begin{align}
H_0(t) = \frac{\Omega(t)}{2}[\cos\Phi(t)\sigma_x + \sin\Phi(t)\sigma_y] + \frac{\Delta(t)}{2}\sigma_z,
\end{align}
where $\Omega(t)$ is the driving field (envelope), $\Phi(t)$ is the phase field, and $\Delta(t)$ is the detuning of our control. Henceforth, we omit the explicit time-dependencies of the various fields for notational brevity; we will explicitly state if a quantity is constant. To relate these fields to the space curve, we first differentiate Eq.~\eqref{scqc_main_eq} with respect to time. We find
\begin{align}
    \vec T \cdot \sigvec  = \hat z \cdot \sigvec_{U_0}, \qquad \vec T \equiv \dot{\vec r},\label{scqc:tangent_eq}
\end{align}
where $\sigvec_{U_0} \equiv U_0^\dagger \,\sigvec\, U_0$ is the Pauli vector in the interaction picture, and $\vec{T}$ is the \emph{tangent curve}. From the operator norm, we observe that $\|\vec{T}\|_2 = 1$. This equation signifies that time in the quantum evolution is equivalent to the \textit{arclength} parameterization of the curve \cite{SHIFRINDIFFERENTIALGEOMETRYFirstCourse2024,
WILLMOREIntroductionDifferentialGeometry2013}. 

Differentiating a second time leads to two more equations that relate properties of the space curve to control fields:
\begin{align}
    \dot{\vec T} \cdot \sigvec = -\Omega\,(\sin\Phi\, \hat x - \cos\Phi\, \hat y)\cdot \sigvec_{U_0}  \label{scqc:dot_t_eq},\\
   (\vec T \times  \dot{\vec T}) \cdot \sigvec = -\Omega\,(\cos\Phi\, \hat x+ \sin\Phi\, \hat y  )\cdot \sigvec_{U_0}  \label{scqc:doubledot_t_eq}.
\end{align}
The details are given in Appendix \ref{scqc_app_equations}. From these, we can show
\begin{align}
    \Omega(t) &= \kappa(t) \label{envelope_cond}, \\
    \dot \Phi(t) - \Delta(t) &=\tau(t),
    \label{phi_delta_cond}
\end{align}
where $\kappa$ and $\tau$ are the signed curvature and torsion of the space curve, respectively. Both $\kappa$ and $\tau$ can be expressed in terms of $\vec{T}$ and its derivatives (see Appendix \ref{scqc_app_mvframe_calc}). Geometrically, the curvature quantifies how much the curve bends (deviates from a straight line), and the torsion quantifies how much the curve twists (deviates from a planar curve). 

The relations between curvature/torsion and the control fields, Eqs.~\eqref{envelope_cond} and \eqref{phi_delta_cond}, along with the identification of the arclength as the time variable, constitute the set of elements required to describe the design of DCGs as a curve construction problem. This reformulation allows us to sidestep the integration of the Schr\"{o}dinger equation and significantly simplify the control design procedure.

In order to complete the description of SCQC, it is necessary to explain how a gate is fixed within this framework. We can use Eqs.~\eqref{scqc:tangent_eq}--\eqref{scqc:doubledot_t_eq} to relate $U_0$ to the space curve derivatives at any given time \cite{ZENGGeometricFormalismConstructingArbitrary2019,DONGDoublyGeometricQuantumControl2021}. This is described in detail in the next section.

\section{Automating the robust curve design procedure}
\label{barq_prelimis_method}

In this section, we describe the key result of this paper, our method called the Bézier Ansatz for Robust Quantum (BARQ) control. We will introduce an alternative description of SCQC in terms of the adjoint representation of the associated unitary and derive the equivalent SU(2) gate-fixing constraints. By utilizing a particular feature of Eq.~\eqref{phi_delta_cond}, we design gates that have unit fidelity in the noise-free case by construction, allowing ample flexibility for the optimization of robustness properties and pulse characteristics without any tradeoff. With the particular choice of the Bézier curve ansatz, we enforce \emph{a priori} curve constraints that guarantee a dephasing-robust evolution generated by a pulse whose envelope vanishes at the initial and final times---these features are built-in upfront without optimization. The numerical optimizer is only needed to cancel additional noise errors or to incorporate additional pulse-shape constraints. Our implementation of SCQC and BARQ is available in the \texttt{qurveros} Python package.

\subsection{Preliminaries}

In this section, we introduce several important concepts that are crucial to BARQ. These include the Frenet-Serret frame, the adjoint representation formulation of SCQC, and a trick we call \emph{total torsion compensation}. Together, these three ingredients allow us to fix the ideal target gate outright by imposing boundary conditions on the space curve, instead of having to introduce a separate gate-fixing term in the cost function used to optimize the space curve shape. We also introduce the idea of Bézier curves, which are fundamental to how BARQ performs space curve optimization.

\subsubsection{Frenet-Serret frame, adjoint representation, and gate-fixing}
During the design procedure, every DCG algorithm deals with some combination of quantities that are local or global with respect to time. For example, while control fields themselves can be viewed as time-local quantities, the evolution they generate is generally viewed as global (i.e., time-non-local) because it depends on the time-ordered exponentiated integral of the Hamiltonian, which involves contributions from all the previous time instants. In principle, both the gate-fixing and noise-robustness goals are global in nature since they both involve integrals over time; however, as we show in this subsection and below in Sec.~\ref{ttc_sec}, gate-fixing can actually be made time-local in SCQC. This stems from the fact that some quantities that appear global become time-local when mapped to space curves. The converse can also happen. Moreover, we have already seen that at least one noise-robustness constraint can also be made time-local in the sense that closure of a space curve is a constraint on the curve's endpoints and not on its overall shape. This conversion from global to time-local constraints is one of the main advantages afforded by SCQC.

To show how gate-fixing can be converted into a time-local constraint within SCQC, it helps to first introduce the Frenet-Serret frame and to reformulate SCQC in the adjoint representation. At each point along a space curve, we can assign an orthonormal basis given by the tangent, normal and binormal vectors, denoted as $\vec T,\vec N,\vec B$ respectively. The normal vector is defined as
$\vec N = \dot{\vec T}/\kappa(t)$, while the binormal is $\vec B  = \vec T \times \vec N$, so that these vectors follow a right-hand rule. Together, they form an orthonormal frame at each point along the curve called the Frenet-Serret (FS) frame, and the orientation of this frame rotates as it moves along the curve. Note that extra care is needed to define this frame at inflection points, i.e., points where the driving field envelope, or equivalently the curvature, vanishes. In this work, we deviate slightly from the standard definition of the FS frame and instead work with a \emph{continuous} FS frame, which is made possible by working with a signed curvature, i.e., the curvature can assume both positive and negative values (see Appendix \ref{scqc_app_mvframe_calc}). The use of a continuous frame allows us to directly relate the FS vectors to the evolution operator, as we now discuss.

From Eqs.~\eqref{scqc:tangent_eq}--\eqref{scqc:doubledot_t_eq}, it appears that the initial orientation of the frame is fixed by the initial condition for $U_0$, $U_0(0)=I_2$, where $I_2$ is the identity operator of the qubit. However, this need not be the case, and in order to keep the gate design purely in the design space of the curve, we reformulate these equations for an arbitrary initial condition. In this regard, for a propagator $U_F$ generated by the same Hamiltonian $H_0$ but with arbitrary initial condition $U_F(0)$, we rewrite the equations using the FS frame vectors as
\begin{align}
-\vec B \cdot \sigvec &= ( \cos\Phi \hat x + \sin\Phi \hat y)\cdot \sigvec_{U_F}, \label{preadj_B} \\
\vec N \cdot \sigvec &=  (-\sin\Phi\hat x +\cos\Phi \hat y)\cdot \sigvec_{U_F},\label{preadj_N}\\
\vec T \cdot \sigvec &= \hat z \cdot \sigvec_{U_F}.\label{preadj_T}
\end{align}
The initial condition $U_F(0)$ is set by the initial orientation of the frame and by $\Phi(0)$.
\begin{figure}[htbp]
\centering
    \includegraphics[trim = {1in 7.1in 4in 0.75in},
    clip]{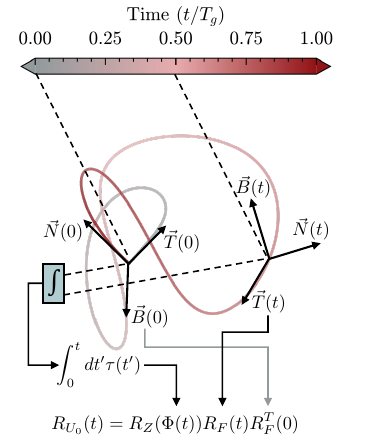}
    \caption{ Quantum evolution as a space curve. In SCQC, time is the arclength parameterization of the curve: As time evolves, the curve is traversed, starting from the origin at $t=0$ and ending at a final point at $t=T_g$. The evolution operator in the adjoint representation at time $t$, $R_{U_0}(t)$, is obtained from the space curve by calculating all the inner products between the Frenet-Serret frame vectors at $t=0$ and at time $t$ and by calculating the integral of the torsion up to time $t$.}
    \label{adjoint_rep_scqc_illu}
\end{figure}

Next, we introduce the adjoint representation \cite{BYRDBangBangOperationsGeometric2002} (also known as the control matrix \cite{GREENArbitraryQuantumControlQubits2013}) to streamline the process of gate-fixing in SCQC. By definition, for a unitary $U$, the elements of the adjoint representation of the evolution are given by
\begin{align}
    R_U^{ij} = \frac{1}{2}\text{tr}(U^\dagger\sigma_iU\sigma_j).
\end{align}
Using this definition and examining Eqs.~\eqref{preadj_B}--\eqref{preadj_T}, we can express the adjoint representation of $U_F$ as
\begin{align}
    R_{U_F}(t) = R_Z(\Phi(t))R_F(t),
\end{align}
where the rows of matrix $R_F$ contain the FS vectors:
\begin{align}
    R_{F}(t) = \begin{bmatrix}
        -\vec B\hphantom{-} \\
        \hphantom{-}\vec N\hphantom{-} \\
        \hphantom{-}\vec T \hphantom{-}
    \end{bmatrix},
    \label{rotmatF_def}
\end{align}
and 
\begin{align}
    R_Z(\Phi(t)) =  \begin{bmatrix}
        \cos\Phi(t) & -\sin\Phi(t) & 0 \\
        \sin\Phi(t) & \hphantom{-}\cos\Phi(t) &0 \\
        0&0&1
    \end{bmatrix}.
\end{align}
Since $U_0(0)=I_2$, we have $U_0(t) = U_F(t)U_{F}^\dagger(0)$, which in the adjoint representation yields
\begin{align}
    R_{U_0}(t) = R_Z(\Phi(t))R_F(t) R_F^T(0).
    \label{adjoint_rep_scqc}
    \end{align}
In Fig. \ref{adjoint_rep_scqc_illu}, we provide an illustration of Eq. \eqref{adjoint_rep_scqc}. Except for the $Z$-rotation, which depends on a global quantity, all elements of the adjoint representation evolution are found by taking inner products of the FS frame vectors at two different times. The reason $R_Z(\Phi(t))$ is global is because $\Phi(t)$ itself is time-non-local from the point of view of the space curve. This is because it is related to the time-integral of the torsion, i.e., \emph{total torsion}, as follows from Eq.~\eqref{phi_delta_cond}. This is an example in which a quantity that is normally viewed as time-local---a control field---becomes global in SCQC.

From Eq.~\eqref{adjoint_rep_scqc}, we see that to fix the final gate, we need to choose the relative orientation of the initial and final FS frame and choose the value of $\Phi(T_g)$ appropriately. Thus, two of the three real parameters that specify a single-qubit unitary are fixed by the boundary conditions of the FS frame and are thus local in SCQC, while the third parameter is fixed by the geometrically global parameter $\Phi(T_g)$.
The fact that most of the gate-fixing constraints are encoded in local properties of the frame is key to the BARQ method, as we discuss in Sec.~\ref{sec:barq}. In Sec.~\ref{ttc_sec}, we show that $\Phi(T_g)$ can be fixed without having to impose a time-non-local constraint on the space curve, so that gate-fixing is a fully local procedure in our formulation of SCQC.

Overall, using SCQC to associate the control fields with geometric elements of a space curve, we not only avoid the integration of the Schr\"{o}dinger equation (which often requires numerical methods) but we can also separate the task of gate-fixing into local and global elements. Had we started with an ansatz for the control fields, the gate-fixing problem typically involves a coupled set of differential equations, which can either obscure the local elements of the design or even eliminate them.

Going beyond unitary evolutions, similar geometric meanings can be given in the context of quantum channels and one of their representations, namely the transfer matrix or superoperator formalism \cite{CHOWUniversalQuantumGateSet2012,
GILCHRISTVectorizationQuantumOperationsIts2009,
GREENBAUMIntroductionQuantumGateSet2015,
GYAMFIFundamentalsQuantumMechanicsLiouville2020,
HANGLEITERFilterfunctionFormalismSoftwarePackage2021,
NIELSENGateSetTomography2021,
HASHIMPracticalIntroductionBenchmarkingCharacterization2024}. Although in the main text we restrict ourselves to the Pauli algebra, by using the Liouville representation (see Appendix \ref{liouville_scqc}), we lay the foundation for a systematic procedure to extend the geometric meaning of SCQC to more general contexts.

\subsubsection{Removing global constraints with total torsion compensation}
\label{ttc_sec}

In the previous section, we showed that gate-fixing in SCQC nominally involves a combination of local and global space curve constraints. The only global constraint comes from the $Z$-rotation angle $\Phi(T_g)$, which is related to the total torsion of the space curve, as discussed above. We can decompose the target gate in the adjoint representation, $R_g$, as $R_g = R_Z(\theta) R_g'$ for some angle $\theta$. If we then choose the initial and final FS frame vectors such that
\begin{align}
    R_F(T_g) R_F^T(0) = R_g',
\end{align}
then we must additionally impose the global constraint 
\begin{align}
    \Phi(T_g) = \theta + 2k\pi, k\in \mathbb{Z},
    \label{fidelity_z_rot}
\end{align}
in which case the target gate is reached with unit fidelity. Note that, in the adjoint representation, the gate fidelity is given by (see Appendix~\ref{liouville_scqc})
\begin{align}
   \mathcal{F}_g(U_0(T_g)) = \frac{d+1+\text{tr}(R_g^TR_{U_0}(T_g))}{d(d+1)},
\end{align}
where $d=2$ is the Hilbert space dimension.
An obvious way to impose the above constraint on $\Phi(T_g)$ would be to include it as a term in the cost function for the space curve optimizer, but this would then lead to an undesirable tradeoff between gate-fixing and noise-robustness.

One way to avoid the global constraint would be to implement the final $Z$-rotation virtually~\cite{MCKAYEfficientGatesQuantumComputing2017}. Alternatively, we can instead employ a trick that we call \emph{total torsion compensation} (TTC). The starting point for TTC is to notice that there is a type of gauge invariance in the mapping that relates the curvature and torsion of a space curve to the control fields, specifically Eq.~\ref{phi_delta_cond}. From this equation, it is evident that we can add any function of time $w(t)$ to both $\dot\Phi$ and $\Delta$ without changing the torsion. We can interpret this gauge freedom as the statement that a given space curve maps to a family of control Hamiltonians related by $\Phi(t)\to\Phi(t)+\int_0^t dt'w(t')$, $\Delta(t)\to\Delta(t)+w(t)$. This means that we can always attain the target value of $\Phi(T_g)$ regardless of what the total torsion is, provided we choose the detuning $\Delta$ such that $\Phi(T_g)-T_g\Delta$ is equal to the total torsion $\int_0^{T_g}dt\tau(t)$. That is, given the $Z$-rotation angle $\theta$ in the target gate, we choose the detuning according to
\begin{align}
    T_g \Delta =  \theta + 2k\pi -\int_0^{T_g}\,dt \tau(t) , k\in \mathbb{Z}\label{ttc_eq}, 
\end{align}
in which case Eq.~\eqref{fidelity_z_rot} is automatically satisfied \emph{without} imposing a global constraint on the space curve. In this way, we can design a space curve without constraining the total torsion and afterwards use TTC, Eq.~\eqref{ttc_eq}, to ensure the target gate is achieved with unit fidelity in the absence of noise. Any deviation of the total torsion away from the target value $\Phi(T_g)$ is effectively absorbed into the detuning. In the present work, we use the value of $k$ that minimizes the magnitude of the detuning, but any choice of the integer $k$ accomplishes the same goal. Note that TTC can be regarded as equivalent to the phase-ramping method discussed in Refs.~\cite{GAMBETTAAnalyticControlMethodsHighfidelity2011,PATTSingleMultiplefrequencyshiftedLaminarPulses1992}. TTC is one of the key elements of the BARQ method, as it enables the optimization step of BARQ to focus solely on the noise-robustness properties of the quantum evolution by eliminating the global gate-fixing constraint.

Note that in cases where it is not possible to apply a constant nonzero detuning, we can still resort to optimizing the space curve shape to account for this residual $Z$-rotation in the target gate, either as part of the main optimization step of BARQ or in a post-optimization manner, by optimizing a subset of parameters. This separation of optimization tasks can still alleviate the tradeoff between gate-fixing and pulse-robustness when the detuning cannot be chosen to satisfy Eq.~\eqref{ttc_eq}.

\subsubsection{B{\'e}zier curves}

The last element that enables BARQ to systematically create robust quantum control fields is the usage of the B{\'e}zier curves. These curves have been famously used in the field of Computer Aided Geometric Design  \cite{FARINCurvesSurfacesCAGDPractical2001}.

We begin the analysis with the position vector of the B{\'e}zier curve~\cite{ERKANSerretFrenetFrameCurvaturesBezier2018}
\begin{align}
    \vec r(x) = \sum_{j=0}^{n} \vec w_j g_{j,n}(x),\qquad x \in [0,1], \label{bezier_curve_eq}
\end{align}
where the curve is defined through $n+1$ control points $\vec w_j$, and the interpolant functions are the Bernstein polynomials given by
\begin{align}
    g_{j,n}(x) = \begin{pmatrix}
        n \\ j
    \end{pmatrix} x^j(1- x)^{n-j}.
\end{align}
An important property of the B{\'e}zier curves is that they transfer the parameter differentiation to the control points~\cite{FARINCurvesSurfacesCAGDPractical2001}:
\begin{align}
  \frac{d^q}{dx^q}  \vec r (x) = \frac{n!}{(n-q)!} \sum_{j=0}^{n-q} \Delta^q \vec w_j g_{j,n-q}(x),
  \label{Bézier_diff}
\end{align}
where $\Delta^q$ is the $q$-th order forward difference operator, defined recursively as $\Delta^q = \Delta^{q-1}(\Delta),\, q>1$ and which acts on a control point as $\Delta \vec w_j = \vec w_{j+1} - \vec w_{j}$. Note that the parameter $x$ is not necessarily  the curve's arclength parameterization. 

Another important aspect of the  B{\'e}zier curve ansatz is that the geometric properties of the curve are encoded in the control points. When constraints are imposed on the curve and its derivatives at the endpoints of the curve, only a few control points are affected, since from the properties of the Bernstein basis, it holds that
\begin{align}
    g_{j,n}(0) = \delta_{j0},\qquad g_{j,n}(1) = \delta_{jn} .
    \label{bernstein_boundary_eq}
\end{align}

Under the SCQC formalism, a first-order dephasing robust quantum evolution is associated with a closed curve. Although such a condition is global in the time domain, it becomes a local property in the curve domain. Evaluating the Bézier curve at the endpoints using Eq.~\eqref{bernstein_boundary_eq}, we find that only the vectors $\vec w_0$ and $\vec w_n$
are involved. We can therefore force the curve to be closed simply by setting
\begin{align}
    \vec w_0 = \vec w_{n} = \vec 0.
    \label{closed_curve_control_points}
\end{align}
We further impose the constraint for the driving field to vanish at the boundaries of the evolution. This requirement often enters into the design as an experimental constraint and is also sometimes necessary for methods that utilize envelope-dependent frame transformations, for instance to suppress leakage~\cite{GAMBETTAAnalyticControlMethodsHighfidelity2011}. In terms of the Bézier curve, this requirement translates to the condition that the control points near the endpoints of the curve must be locally co-linear:
\begin{align}
    \vec w_2  &\parallel  \vec w_1, \qquad \text{for}\,\, \kappa(0) = 0, \\
    \vec w_{n-2}  &\parallel  \vec w_{n-1}, \qquad\text{for}\,\,  \kappa(T_g) = 0.
\end{align}

In summary, by enforcing the conditions above, we automatically guarantee that the resulting space curve corresponds to a quantum evolution that is first-order-robust to dephasing noise that is generated by a pulse with vanishing envelope at the initial and final times.

\subsection{The BARQ method}\label{sec:barq}

\begin{figure*}[htbp]
\centering
    \includegraphics[trim = {0.5in 7.4in 0.5in 0.2in},
    clip]{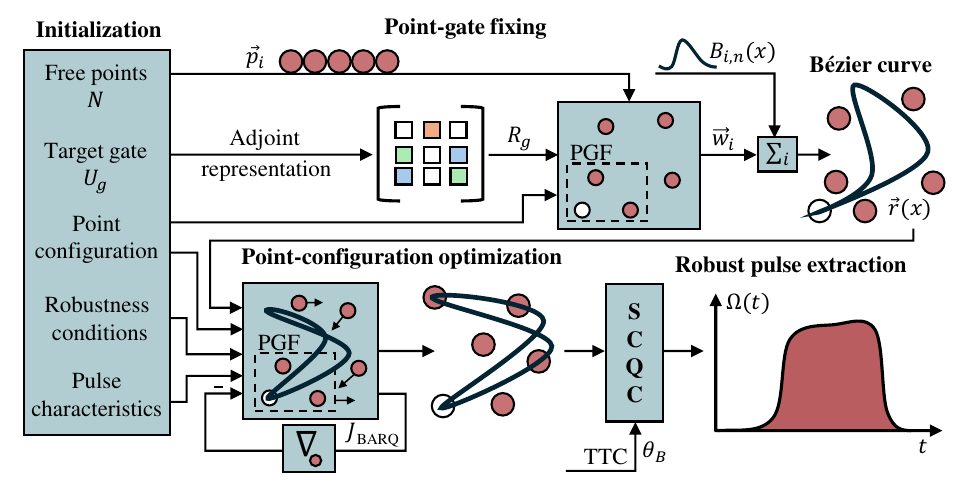}
    \caption{Schematic of the BARQ method. The initial information required for the method is found on the left side of the figure. This includes $N$ free points to optimize over, the target gate $U_g$, and the point configuration (PC). The PC contains both static and optimizable parts. The robustness conditions and the pulse characteristics are included as terms in the total objective function $J_{\text{BARQ}}$ with associated weights. The top half of the figure illustrates the gate-fixing procedure. The adjoint representation of $U_g$ is calculated and then encoded in the space curve by constraining a few control points near the endpoints of the curve using the PGF part of the PC. The Bézier curve is constructed using the control points $\vec w_i$ and the Bernstein basis using Eq.~\eqref{bezier_curve_eq}. In the bottom half of the figure, the remaining set of control points are optimized, while the constraints that encode $U_g$ remain intact throughout the optimization.
    Using TTC, we can achieve unit fidelity in the noise-free case. Following the SCQC framework, the robust pulses are extracted from the optimized space curve by mapping geometric quantities to the control fields.
    \label{barq_illu_fig}}
\end{figure*}

Equipped with the preliminaries described in the previous subsection, we can now introduce the BARQ method. We begin by imposing the control point constraints derived above that guarantee upfront that the Bézier curve is closed and has vanishing curvature at the endpoints. In the remainder of this section, we elaborate on how to additionally set the target gate by fixing a subset of the control points appropriately and on how we optimize over the remaining control points to achieve further desired properties in the corresponding quantum evolution. Figure~\ref{barq_illu_fig} summarizes the workflow of BARQ.

\subsubsection{Fixing the target gate in BARQ}

Fixing the target gate amounts to enforcing that the final noise-free evolution operator in the adjoint representation, $R_{U_0}(T_g)$, equals the target gate $R_g$. As mentioned above, extra care is needed to define the FS frame at inflection points (points where $\kappa=0$). As we discuss in Appendix~\ref{scqc_app_mvframe_calc}, if $\dot{\vec{T}}$ changes sign at an inflection point, then the sign of $\kappa(t)$ must also change in order to ensure that the FS frame, and hence the evolution operator, remains continuous at that point. We refer to these inflection points as \emph{singular} points. As discussed in Appendix \ref{Bézier_curve_calc}, singular points can lead to additional minus signs in the FS frame at the final time $T_g$. Because the number of singular points can vary during numerical curve optimization, we must modify the gate-fixing procedure slightly to account for these possible sign corrections in the final FS frame. In the case of Bézier curves that contain $M$ singular points, Eq.~\eqref{adjoint_rep_scqc} becomes
\begin{align}\label{eq:adjoint_evolution_with_M}
    R_{U_0}(T_g) = R_Z(\Phi(T_g)+(M+1)\pi)R_B(T_g)R_B^T(0),
\end{align}
where $R_B$ is the matrix of final FS frame vectors up to factors of $(-1)^{M+1}$ caused by the singular points (see Appendix \ref{Bézier_curve_calc}). We have effectively moved these factors into the $Z$-rotation in Eq.~\eqref{eq:adjoint_evolution_with_M}, and in so doing we can now fix $R_B(T_g)$ completely by choosing the final control points appropriately. To fix the gate, we require
\begin{align}
    R_B(T_g) = R_Z^T(\Phi(T_g)+(M+1)\pi)R_gR_B(0).
\end{align}
The control points in $R_B(0)$ and $R_B(T_g)$ contain only local information so we will first require
\begin{align}
    R_B(T_g) = R_Z^T(\theta_B)R_gR_B(0),
    \label{gen_barq_points}
\end{align}
where $\theta_B$ is a free parameter that we refer to as the BARQ angle. Defining $\vec{a}_i$ to be the rows of the matrix $R_gR_B(0)$ and choosing a value for $\theta_B$, we can satisfy Eq.~\eqref{gen_barq_points} by setting
\begin{align}
    \vec w_{n-1} &= - \lambda_{n-1}^+ \,\vec{a}_3, \label{gatefix_wn1}\\
    \vec w_{n-3} &= \lambda_{n-3}^+(\sin\theta_B\, \vec{a}_1 -\cos\theta_B\, \vec{a}_2) - \lambda_{n-3}\,\vec{a}_3,\label{gatefix_wn3}
\end{align}
where the $\lambda_i$ are additional free parameters associated with the $i$-th control point, and the + superscript indicates that the respective quantity must be chosen to be positive. Parameters with the same subscript but different superscripts are distinct.

By adopting this choice of control points, the average gate fidelity is now
\begin{align}
    \mathcal{F}_g(U_0(T_g)) = \frac{2+\cos(\Phi(T_g)+(M+1)\pi-\theta_B)}{3}
    \label{fidelity_ang_diff}
\end{align}
In order to fix the gate exactly with unit fidelity, we can utilize the TTC discussed earlier in Sec.~\ref{ttc_sec} after the numerical curve optimization is completed and both $M$ and the total torsion are known. Here, we need to modify the TTC slightly to account for not only the total torsion but also the signs induced by the singular points:
\begin{align}
    T_g \Delta &=  \theta_B + (2k^*-M-1)\pi -\int_0^{T_g}\,dt \tau(t) \label{ttc_barq},\\
    k^* &= \underset{k}{\text{argmin}} |T_g\Delta|.
\end{align}

With this set of equations, we are able to encode the target gate in the control points of the B{\'e}zier curve and achieve unit fidelity \emph{a priori}, provided that the TTC is employed. We have also implicitly assumed that $\vec w_1 \nparallel \vec w_3$, which is always possible since these are free parameters.

\subsubsection{The optimization scheme}

Figure~\ref{barq_illu_fig} illustrates the BARQ workflow.
Having encoded the target gate and ensured the curve is closed with vanishing curvature at the endpoints, we can now focus solely on the optimizing the remaining control points to build other robustness properties into the quantum evolution, beyond first-order robustness against dephasing noise. At this stage, we can also include pulse characteristics in the cost function to ensure experimentally feasible control fields. 

In BARQ, each additional desired robustness and pulse property beyond the automatic ones described above is achieved by minimizing an objective or loss function $J$ with respect to the control points that have not yet been fixed. All objectives are combined into one total objective $J_{\text{BARQ}}$, which is a linear combination of the individual loss functions with adjustable weights. While all control points participate in the optimization, every objective is affected differently by each point. For instance, if we require the total (oriented) area of the curve to vanish, a condition that achieves second-order static dephasing robustness~\cite{ZENGGeometricFormalismConstructingArbitrary2019}, the control points away from the endpoints of the curve will have a stronger impact on the minimization of the objective. While the control points can be optimized freely, enforcing additional relationships between them can benefit the optimization stage. We refer to the set of choices that enforce any arrangement of control points, and additional relationships between them, as a Point Configuration (PC).

In this context, Eq.~\eqref{closed_curve_control_points}, which enforces the closed curve condition, can be regarded as part of the PC. The choices included in the PC are instrumental to designing robust pulses with desired properties. A choice in the PC can affect either an individual control point or establish a relationship between control points. At a minimum, the PC comprises $N$ free points $\vec p_i$, which are converted to control points $\vec w_i$, through suitable transformations that can help the optimization become more controllable. We found empirically that a useful PC choice for BARQ is to make the normalized orientations and overall scale factors in the first three control points independent parameters:
\begin{align}
    \vec w_1 &= \lambda_1^+ \hat{p}_1 \label{gf1},\\
    \vec w_2 &= \lambda_2 \hat{p}_1 \label{gf2},\\
    \vec w_3 &= \lambda_3^+\, \hat p_2 + \lambda_3 \,\hat p_1\label{gf3}.
\end{align}
This choice decouples the scale of the free points from their direction and provides more control over the endpoints of the curve, which can be used to control the rise-time of the pulse. The same free points participate in the gate-fixing procedure by forming $R_B(0)$. In this regard, this particular choice in the PC encodes the target gate.

In order to set the stage for a more fine-grained approach to the curve optimization problem, we define two separate control point categories within the PC. The points that participate in fixing the target gate and lie near the endpoints of the curve constitute the point-gate-fixing (PGF) category. Beyond encoding the target gate, the PGF is able to control the early- and late-time behavior of the pulse, while it has a significantly lesser impact on other objectives that depend on global features of the curve. On the other hand, the remainder of the points that lie away from the endpoints belong to the point-robustness structure (PRS) category, which primarily handles the robustness properties that are expressed in terms of global quantities, but can also be used to control characteristics of the pulse in the intermediate times of the evolution. For instance, if multiplicative driving field error needs to be suppressed, the area traced by the tangent must vanish in three orthogonal directions~\cite{NELSONDesigningDynamicallyCorrectedGates2023}. This condition is expressed as a global quantity, so the PRS will exercise the maximum influence on it.

In essence, the PC contains both fixed and optimizable parts. The simplest form of optimization is to leave the PGF category static and optimize the elements of the PRS category, by setting the control points equal to the free points except for the ones used in the PGF. More sophisticated PC choices can exhibit better adaptation of a pulse in a realistic experimental setting, while the gate fidelity remains unchanged throughout the optimization. Given the highly symmetric character of the vanishing tangent area condition, we believe that a properly defined PRS could encode a discrete symmetry such that the curve ansatz achieves this property without any optimization, similar to the closed curve condition in Eq.~\eqref{closed_curve_control_points}. We have left this investigation to follow-up work.

\subsection{The qurveros software package}

In this section, we briefly describe the provided software and the workflow to design and extract robust pulses using \texttt{qurveros} \footnote{The url of the repository is: \url{https://github.com/evpiliouras/qurveros}}. For a more comprehensive introduction, we refer the reader to the examples folder of the repository.

At the top level, the user is provided with the \texttt{SpaceCurve} class. The class encapsulates all the important information and functionality to establish the curve-to-quantum-evolution mapping, evaluate the robustness conditions, and extract the pulses under a desired control configuration. For a general space curve, the user defines a function \texttt{curve} with signature \texttt{curve(x, params)}. The first argument refers to the parameter used to traverse the curve (not necessarily the arclength) and the second argument contains the auxiliary parameters that control the characteristics of the curve. In this context, the BARQ parameters belong to the set of auxiliary parameters \texttt{params}, while the argument \texttt{x} is the variable $x$ that is used for the Bernstein basis.

The provided software is built using JAX~\cite{BRADBURYJAXComposableTransformationsPython+NumPy2018} at its core, where automatic differentiation allows efficient calculation of curve derivatives and parameter gradients. Using primarily JAX's NumPy API, the curve function is defined with the following piece of code as follows:
\begin{verbatim}
def curve(x, params):
        
        x_comp = f(x, params)
        y_comp = g(x, params)
        z_comp = h(x, params)

        return [x_comp, y_comp, z_comp]
\end{verbatim}
where the argument \texttt{x} stands for the traversal parameter and \texttt{params} are auxiliary parameters that can control various curve properties.
Having defined the function object, an instance is created with
\begin{verbatim}
spacecurve = SpaceCurve(curve=curve,
                        order=order, 
                        interval=[x_0, x_1],
                        params=params)
\end{verbatim}
where the argument \texttt{order} defines whether the curve function refers to the position vector or the tangent vector. The \texttt{interval} defines the range of the traversal parameter, and the \texttt{params} are the values of the auxiliary parameters, if required for the curve function.

All the geometric quantities that are used in the SCQC formalism are stored in the \texttt{frenet\_dict} attribute. In order to evaluate the robustness properties of the curve, we use
\begin{verbatim}
spacecurve.evaluate_frenet_dict()
spacecurve.evaluate_robustness_properties()
\end{verbatim}
The extraction of the control fields is implemented with an instance method. Considering a control Hamiltonian of the form
\begin{align}
    H_0 = \frac{1}{2}\sum_{i=x,y,z} \Omega_i\sigma_i,
\end{align}
the method is invoked with a string argument that determines which axes of control are nonzero, using the command
\begin{verbatim}
spacecurve.evaluate_control_dict(control_mode)
\end{verbatim}
For example, resonant control is recovered when \texttt{control\_mode="XY"} and when the control uses TTC, we set \texttt{control\_mode="TTC"}. The developed software provides the subpackage \texttt{qubit\_bench} that automates the noise simulations used in this work, including the ones necessary for the validation of the evolution's robustness properties.

To facilitate the optimization, we utilize the derived class \texttt{OptimizableSpaceCurve}, where the default parameters are optimized with respect to a loss function which is constructed by providing lists of the form \texttt{[function(frenet\_dict), weight]}. The function is evaluated based on some entries of the \texttt{frenet\_dict}, and the total loss is constructed as a linear combination using the supplied weights. We will explain its usage by introducing the optimization stage of BARQ.

To implement BARQ, we create an instance of the class  \texttt{BarqCurve}, using the command
\begin{verbatim}
barqcurve = BarqCurve(adj_target=adj_target,
                      n_free_points=N,
                      pgf_mod=PGF,
                      prs_fun=PRS)
\end{verbatim}
The first argument refers to the adjoint representation of the target operation, while the second dictates the number of free points used in BARQ. The remaining two refer to particular choices for the PGF and the PRS, supplied as functions. When the PGF and the PRS are not defined, the PGF encodes the target gate with unit scale parameters (with $\lambda_i^+=1,\lambda_i=0$), while the PRS simply acts as an identity operator on the free points. In order to use BARQ to achieve particular robustness properties and pulse characteristics, we invoke the method that prepares the optimization loss as
\begin{verbatim}
barqcurve.prepare_optimization_loss(
        [loss_1(frenet_dict), weight_1],
        [loss_2(frenet_dict), weight_2],
        ...
        [loss_n(frenet_dict), weight_n]
        )
\end{verbatim}
If no particular PC choices are made, we initialize the optimization using
\begin{verbatim}
barqcurve.initialize_parameters(
    init_free_points=init_free_points)
\end{verbatim}
Alternatively, a seed is provided and random points are drawn for the optimization to begin.

The optimization of the internal points is done using Optax~\cite{DEEPMINDDeepMindJAXEcosystem2020}, which supplies the \texttt{optimizer} argument. The optimization begins using the command
\begin{verbatim}
barqcurve.optimize(optimizer, max_iter) 
\end{verbatim}
where the value of \texttt{max\_iter} sets the total number of iterations.

The quantum calculations and simulations described in the following section were performed using QuTiP~\cite{LAMBERTQuTiP5QuantumToolbox2024,ROBERTJOHANSSONQutipQutipQuTiP5102024}. Additional calculations and data analysis were performed using NumPy~\cite{HARRISArrayProgrammingNumPy2020}, SciPy~\cite{VIRTANENSciPy10FundamentalAlgorithms2020} and Pandas~\cite{MCKINNEYDataStructuresStatisticalComputing2010, THEPANDASDEVELOPMENTTEAMPandasdevPandasPandas2024}. The plots were obtained with Matplotlib~\cite{HUNTERMatplotlib2DGraphicsEnvironment2007} using the plotting techniques in Ref.~\cite{VANDERPLASPythonDataScienceHandbook2016} and the settings from Ref.~\cite{SILVAPhysrev_mplstyle2023}.
The software also interfaces with the \texttt{filter\_functions} package, to calculate filter functions and evaluate the gate infidelity under a prescribed time-dependent noise power spectral density~\cite{HANGLEITERFilterfunctionFormalismSoftwarePackage2021,
CERFONTAINEFilterFunctionsQuantumProcesses2021, HANGLEITERFilter_functions2024}.

\section{Results}
\label{results}

In this section, we will assess the performance of BARQ-optimized pulses against quasi-static additive dephasing noise and multiplicative driving field noise. We additionally run simulations with time-dependent dephasing noise and examine the behavior of the fidelity filter functions. 

\subsection{Curve design choices}
As elucidated in the previous section, BARQ provides a tailored ansatz in which first-order dephasing robustness, unit gate fidelity in the noise-free case, and vanishing envelope at the initial and final times of the evolution are all automatically built-in. Additional robustness properties and pulse characteristics can be achieved by optimizing the free control points of the Bézier curve and optionally enforcing custom PGF and PRS configurations.

For the quasi-static noise simulations, we consider a noise Hamiltonian of the form:
\begin{align}
    H_{\text{n}} = \frac{\varepsilon\, \Omega}{2}[\cos\Phi\sigma_x + \sin\Phi\sigma_y] +\frac{\delta_z}{2}\sigma_z,
    \label{static_noise_hamiltonian}
\end{align}
where both errors $\varepsilon, \delta_z$ are assumed to be static throughout the quantum evolution. From a hardware-agnostic point of view, such a noise model arises when an inaccurate calibration is made for the Rabi strength \cite{BALLSoftwareToolsQuantumControl2021} or dephasing errors arise from environmental noise, calibration errors in qubit energy splittings, or other from other effects~\cite{ZEUCHExactRotatingWaveApproximation2020}. In an experimental setting, a similar model is employed in the control of singlet-triplet qubits, where Zeeman energy fluctuations enter as static additive errors and multiplicative driving field errors emerge from electrical-noise induced exchange coupling variations \cite{WALELIGNDynamicallyCorrectedGatesSilicon2024}.
Beyond single-qubit control, similar types of models can be derived when a higher-dimensional control problem is reduced to the control of SU(2) dynamics. A typical example is the implementation of motion-insensitive gates in trapped atom systems \cite{DAMMEMotionInsensitiveTimeOptimalControlOptical2024}.

Here, we demonstrate BARQ by using it to design space curves that generate pulses which can suppress leading-order noise contributions to the gate infidelity. Since the curves are \emph{a priori} closed, the first-order contribution of $\delta_z$ is suppressed. In order to cancel the first-order multiplicative driving field error, we recall the vanishing tangent-area condition found in Ref.~\cite{NELSONDesigningDynamicallyCorrectedGates2023}. 
Since the robustness conditions are enforced by minimizing an appropriate objective function, the cost function that encodes the desired robustness property is given by
\begin{align}
    J_\text{drive} =  \Big\|\int_0^{T_g} dt \,\vec T \times \dot{\vec T}\;\Big\|_2^2 \label{multiplicative_loss}.
\end{align}
\begin{figure}[htbp]
\centering
\includegraphics[trim = {0in 0.2in 0.2in 0in},
    clip]{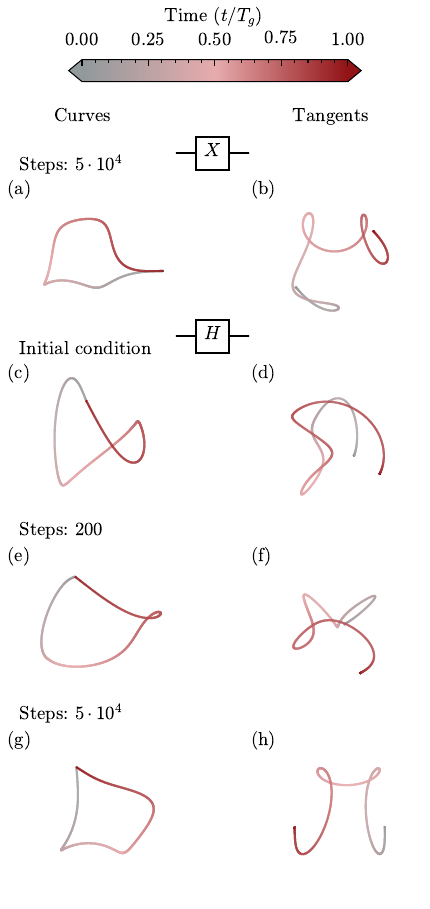}
\caption{BARQ-optimized curves. The left column depicts space curves, while the right column shows the corresponding tangent curves. (a) The space curve and (b) its tangent curve that generate an $X$ gate gate after 50,000 steps of optimization. (c)--(h) The space curve and its tangent for the case of a Hadamard gate at three different stages of the optimization. Each curve is created using 10 free points and optimized according to $J_{\text{BARQ}}$ from Eq.~\eqref{total_cost_eq}. Given the choice of the PGF and the PRS, a set of control points $\vec w_i$ are optimized, and the Bézier curve is constructed using Eq.~\eqref{bezier_curve_eq}, while the tangent curves are found through differentiation with respect to $x$. The total time $T_g$ corresponds to the total length of each curve. 
These highly symmetric shapes reflect their noise-robustness properties.}
    \label{curve_gallery}
\end{figure}
In addition to noise-robustness properties, we will also enforce a pulse characteristic. A simple strategy to reduce the gate time is to increase the peak value of the pulse envelope until the maximum Rabi rate is reached, since the dynamics are equivalent under rescaling~\cite{BERNARDOTimerescaledQuantumDynamicsShortcut2020}. In an effort to achieve minimal gate time, we define the objective:
\begin{align}
    J_{\text{Rabi}} = T_g \max_t\{\kappa(t)\} \label{max_amp_loss}.
\end{align}
As this dimensionless quantity gets minimized, the maximum amplitude of the drive is reached, while the gate time reduces. Since the total gate time is equal to the total length of the curve, we would be tempted to design the shortest possible curves. While this would be a reasonable approach, the curvature must remain bounded \cite{ZENGFastestPulsesThatImplement2018}, otherwise
$J_{\text{Rabi}}$ will increase.

As illustrative examples, we will design an $X$ gate and a Hadamard, which are two commonly used gates for quantum computation. For both gates, we make the same PC choices. The curves are constructed with a total of 10 free points $\vec{p}_i$, $i=1...10$, with a symmetric PGF and an identity PRS. The symmetric PGF forces the endpoints to lie near a sphere, by setting all the parameters $\lambda_i^+$ and $\lambda_i$ from Eqs.~\eqref{gatefix_wn1}, \eqref{gf1}, and \eqref{gf2} equal to the same value $\nu >0$, which remains fixed throughout the optimization. This geometric constraint provides an intuitive way to control the rise times of the pulses. For the $X$ gate, we set $\nu_X =0.25$, and for the Hadamard we select $\nu_H = 0.5$. These heuristic choices lead to a satisfactory pulse rise time at the end of the optimization. The values of $\lambda_3$ and $\lambda_{n-3}$ in Eqs.~\eqref{gatefix_wn3} and \eqref{gf3} (which do not change the directions of the binormal vectors at the endpoints but do influence the magnitude of the respective control points) are set equal but still optimizable, so that the optimization is not unnecessarily constrained, while the value of $\lambda_{n-3}^+$ is fixed to $\nu$. As for the PRS (the points controlling the middle of the space curve), we allow these remaining 8 points to move freely throughout the optimization. (The first 2 points are used in Eqs.~\eqref{gf1}--\eqref{gf3}.) The optimization can be run multiple times with different initial values of the free points until a satisfactory curve is obtained.

For this optimization, we use the following total cost function:
\begin{align}
    J_{\text{BARQ}} = J_\text{drive} + 10^{-2}J_{\text{Rabi}},
    \label{total_cost_eq}
\end{align}
and we run the Adam optimizer for 50,000 steps~\cite{KINGMAAdamMethodStochasticOptimization2017}. 
The gradients of Eq.~\eqref{total_cost_eq} are obtained through automatic differentiation using JAX. 

An advantage of using SCQC is that the parameter gradients lie solely in the curve space, in contrast to other methods that may require the integration of the Schr\"{o}dinger equation. In addition, when BARQ is used to optimize the curve ansatz, no gradients of the gate fidelity are required since the gate is automatically fixed through a subset of the control points. From an implementation point of view, we can express all cost functions that use global quantities as nonlinear functions of the control points by using the non-arclength parameter $x$ (see Appendix~\ref{scqc_app_equations}).
\begin{figure*}[htbp]
    \includegraphics{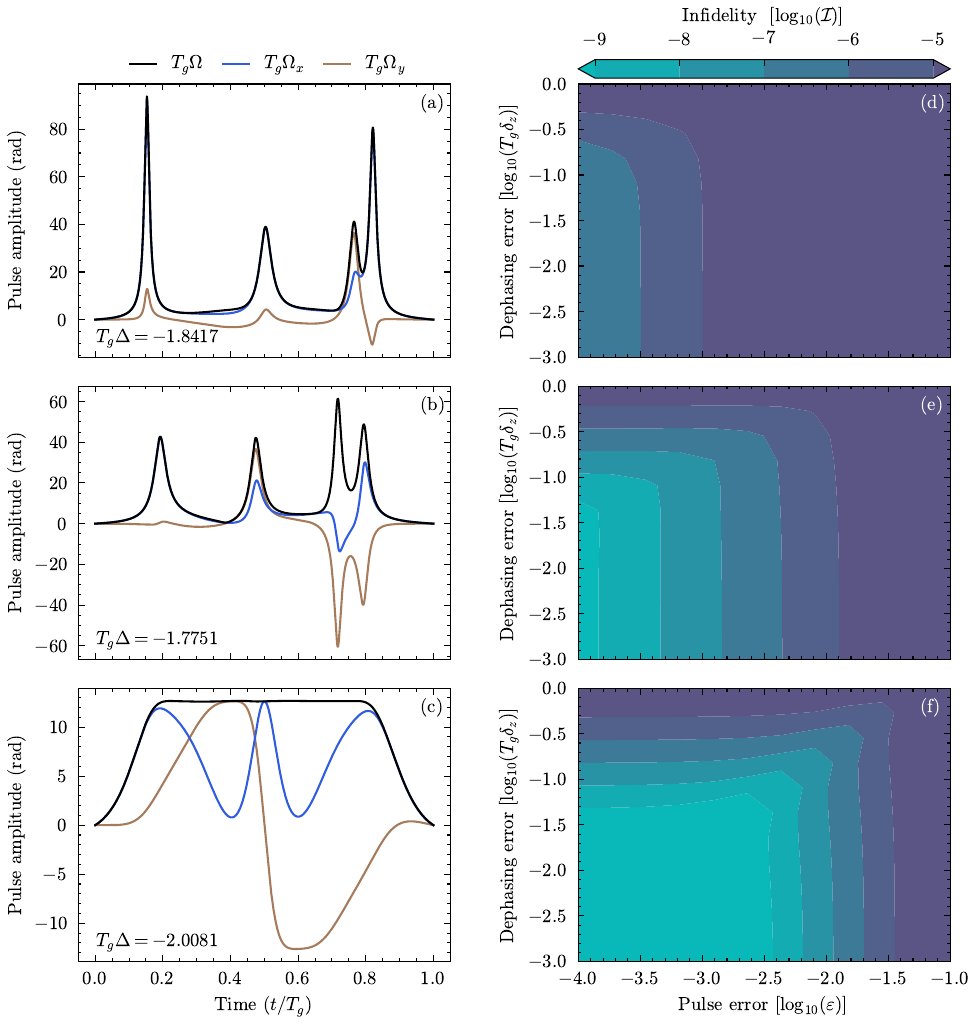}
    \caption{Control fields (a)-(c) and gate infidelities (d)-(f) for the Hadamard gate. Each row depicts a snapshot during the optimization for three different optimization steps ($0,200,5\cdot10^4$).
    While the initial control fields (a) already generate first-order dephasing-robust gates by construction as shown in (d), the initial waveforms are not experimentally friendly.
    As the curve is optimized, multiplicative driving-field error-robustness emerges after 200 steps (e), while the peaks of the control fields become smaller in magnitude (b).
    By the end of the optimization, the region of small infidelity (light blue) has expanded (f), a signature of achieving robustness against both errors to leading-order. In (c), the effectiveness of the $J_{\text{Rabi}}$ constraint in producing smooth pulses is evident.
    For (a)-(c), $\Omega_x = \Omega \cos\Phi$ and $\Omega_y = \Omega\sin\Phi$ in time-normalized units, and the value of the detuning $T_g\Delta$ set by TTC (Eq.~\eqref{ttc_barq}) is shown.}
    \label{hadamard_curve_evol}
\end{figure*}
Figure~\ref{curve_gallery} shows examples of space curves and their tangents generated via this procedure for both the $X$ and Hadamard gates. In the case of the Hadamard gate, we show the curves at three different stages of the optimization: the initial curves, the curves after 200 optimization steps, and the final curves after 50,000 optimization steps. The noise-robustness properties encoded in these curves are immediately evident from their shapes. For example, looking at the final tangent curves shown in Figs.~\ref{curve_gallery}b and \ref{curve_gallery}h, we see that both curves are highly symmetric, which is a natural consequence when the total oriented (projected) area is required to vanish so that driving field error-robustness is achieved.

Moving our attention to the control fields obtained from these space curves, we find that both robust gates are implemented with smooth pulses, as shown in Figs. \ref{hadamard_curve_evol} and \ref{xgate_noise_fields}. The waveforms respect (by construction) the vanishing envelope condition and have small maximal amplitudes as a result of the $J_{\text{Rabi}}$ objective. The value of $\nu$ was tuned heuristically until satisfactory rise times were achieved. While $\nu$ can be regarded as a free parameter, for this work, we keep it fixed in order to simplify the optimization stage. 

In order to better grasp the effect of $J_{\text{Rabi}}$, we study the evolution of the control fields for the Hadamard gate as the optimization progresses in Fig.~\ref{hadamard_curve_evol}. The fields always vanish at the initial and final times, which is a behavior enforced directly at the curve level. In Fig.~\ref{hadamard_curve_evol}a, although the curve corresponds to a dephasing-robust evolution, its sharp peaks make it experimentally infeasible. After 200 steps, the peaks are still present, but are reduced by almost half. By the end of the optimization, the envelope is smoothed out, with a small maximum amplitude.

Before we conclude this section, we would like to emphasize that some robustness properties can be achieved purely from a symmetry-enforcement point of view, similar to the closed curve condition enforced directly from the control points in Eq.~\eqref{closed_curve_control_points}. We expect that by building discrete symmetries into the PRS, we can obtain noise-robustness properties for the continuous-time evolution. This strategy can ease the burden introduced when optimizing for multiple quantities and is left for future investigation.

Also before we move on, we again emphasize that throughout the optimization, the final evolution operator is always fixed to the target gate, whereas the control points are optimized only for the robustness properties and any additional pulse constraints.

\subsection{Quasi-static noise simulations}

In this section, we assess the performance of the generated pulses against quasi-static noise based on the noise model from Eq.~\eqref{static_noise_hamiltonian}. The assessment is done using the gate fidelity: \cite{PEDERSENFidelityQuantumOperations2007,NIELSENSimpleFormulaAverageGate2002}:
\begin{align}
   \mathcal{F}(M,I_d) = \mathcal{F}(M) = \frac{\text{tr}(MM^\dagger)+|\text{tr}(M))|^2}{d(d+1)} ,
\end{align}
for $M = U_0(T_g) U_g^\dagger$. In particular, we examine the infidelity as a function of the noise strengths.
\begin{figure}[htbp]
\centering
\includegraphics{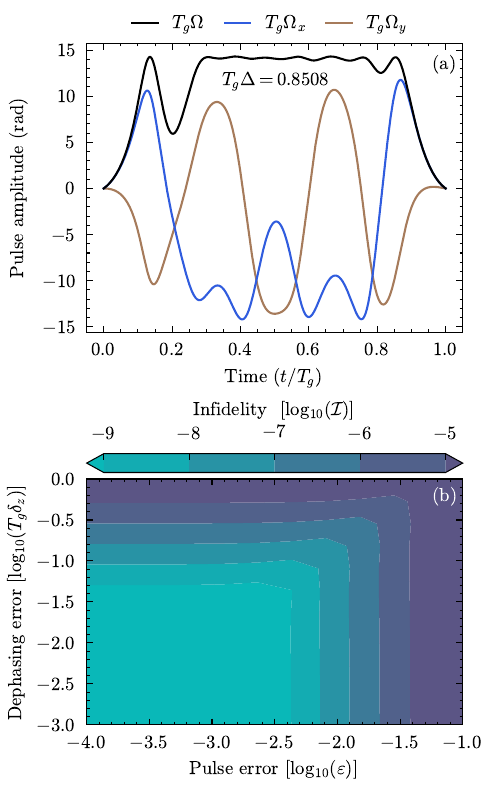}
    \caption{Control fields (a) and gate infidelity (b) for the $X$ gate after 50,000 optimization steps. The gate is designed to be both dephasing- and driving-field error-robust, a feature which is validated in panel (b). The control fields in (a) are smooth with small maximum value and vanishing envelope at the initial and final times, as required by design. In (a), $\Omega_x = \Omega \cos\Phi$ and $\Omega_y = \Omega\sin\Phi$ in time-normalized units, and the value of the detuning $T_g\Delta$ set by TTC (Eq.~\eqref{ttc_barq}) is shown.}
    \label{xgate_noise_fields}
\end{figure}
In the previous section, we laid out the design choices to construct curves that generate robust pulses.
When BARQ is used, the curve ansatz is \emph{a priori} closed (Fig.~\ref{curve_gallery} left column), so that the associated quantum evolution is robust to first-order in $\delta_z$.

This behavior is depicted in Fig. \ref{hadamard_curve_evol}. Starting from Fig. \ref{hadamard_curve_evol}d which is the initial condition of BARQ's Hadamard parameters, we observe two vertical strips that signify dephasing robustness for fixed driving fields errors. After 200 steps, in Fig. \ref{hadamard_curve_evol}d, the increased number of vertical strips demonstrates the pulses' first-order dephasing robustness as driving field error robustness is acquired, with the optimization searching for the appropriate arrangement of the control points at every step.

After the optimization has reached the maximum number of steps, both gates exhibit driving-field error-robustness, as understood from the extension of the light blue region in Figs.~\ref{hadamard_curve_evol}d-\ref{hadamard_curve_evol}f and \ref{xgate_noise_fields}b. and supported by the tangent curve symmetry in Figs.~\ref{curve_gallery}b and \ref{curve_gallery}h.

\subsection{Time-dependent dephasing noise simulations}
In this section, we test the performance of the optimized $X$ and Hadamard gates obtained in the previous sections against time-dependent dephasing noise. We generate wide-sense stationary noise with a power spectral density (PSD) of the form:
\begin{align}
    S(\omega) = \frac{ \lambda^2 T_g}{(\omega/\omega_B)^\alpha}, \qquad \omega_B = \frac{2\pi}{T_g},
    \label{res_noise_psd}
\end{align}
for $\alpha = 1,2$, where $\lambda$ is the noise strength in the appropriate units of its amplitude. Time-dependent dephasing noise with strong low-frequency components is often encountered in the study of solid-state qubits \cite{BEAUDOINMicroscopicModelsChargenoiseinducedDephasing2015}, where dephasing is attributed to charge-noise fluctuations described by a $\omega^{-\beta}$ power-law PSD. This sort of time-dependent dephasing noise can also arise from a phase instability of the local oscillator used to drive the system \cite{BALLRoleMasterClockStability2016}. We consider zero-mean Gaussian-distributed noise that enters into the Hamiltonian through the addition of the following term:
\begin{align}
    H_\text{n} = \frac{\delta_z(t)}{2}\sigma_z,
\end{align}
with variance $\delta_z^2$.
When the noise is stochastic, the assessment of a quantum control strategy requires averaging over noise realizations. Filter functions are often used to facilitate this~ \cite{GREENArbitraryQuantumControlQubits2013,
CERFONTAINEFilterFunctionsQuantumProcesses2021,
HANGLEITERFilterfunctionFormalismSoftwarePackage2021}. In Appendix \ref{adjoint_noise_calc}, we show that the average gate infidelity solely due to the noise (assuming the gate is perfect in the noise-free case) to first order is
\begin{align}
   \mathcal{I}_{z} = \frac{1}{6\pi}\int_{-\infty}^{\infty}d\omega\, S_{z}(\omega)F_{z}(\omega),
   \label{infid_dephasing}
\end{align}
where the index indicates the noise source operator, $S_z(\omega)$ is the PSD of the dephasing noise, and $F_{z}(\omega)$ is the fidelity filter function. In general, the fidelity filter function captures the effect of the control in frequency space with respect to the axis along which the error occurs. In this particular case, the fidelity filter function is given by (see Appendix \ref{cfi_calculation})
\begin{align}
    F_z = \frac{1}{2} \Big\|\int_0^{T_g}dt\, \vec T(t) e^{-i\omega t}\Big\|_2^2.
    \label{dephasing_ff}
\end{align}
Even if the noise PSD is not exactly described by Eq.~\eqref{res_noise_psd}, so long as it can be bounded by a power-law PSD, we can still bound the gate infidelities for a large class of noise spectra.
\begin{figure}[htbp]
\centering
\includegraphics{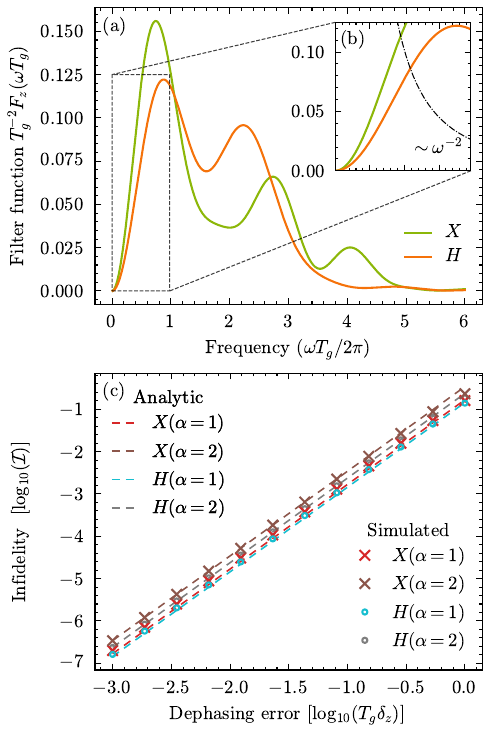}
    \caption{Filter functions and infidelities of the BARQ-optimized $X$ and Hadamard gates in the presence of time-dependent dephasing noise with a $\omega^{-\alpha}$ PSD for $\alpha =1,2$.
    The filter function of the Hadamard gate contains more weight at high frequencies compared to that of the $X$ gate, as expected from its 1.5$\times$ smaller CFI. (b) Low-frequency zoom-in of (a), together with a unit-strength $\omega^{-2}$ PSD with $T_g \omega_B = 1 \,\text{rad}$ (dashed line). The smaller overlap of the Hadamard's filter function with this PSD explains its smaller CFI. (c) The gate infidelity obtained from time-dependent noise simulations for $\alpha=1,2$ averaged over 10,000 noise realizations.
    As expected from (a), (b), for both values of $\alpha$, the Hadamard gate is less sensitive to noise compared to the $X$ gate. The analytic results are derived from the first-order contribution of the cumulant function evaluated at $T_g\delta_z = 1\, \text{rad}$ (see Appendix \ref{adjoint_noise_calc}). The filter functions and the analytic results are performed using the \texttt{filter\_functions} package \cite{HANGLEITERFilter_functions2024}.}
    \label{td_ff_plot}
\end{figure}

In general, the infidelity is calculated through an integration in the frequency domain, as in Eq.~\eqref{infid_dephasing}.
For the particular case where $\alpha=2$, we can execute this calculation in the time domain by exploiting properties of this particular noise power-law. Using an integration by parts in the filter function definition and using the closed-curve condition, the $\omega^2$ term in the denominator is canceled. The infidelity becomes proportional to a geometric quantity that we call the curve filtering index (CFI) (see Appendix \ref{cfi_calculation}):
    \begin{align}\label{eq:cfi}
    \text{CFI} = \frac{1}{T_g^3} \int_0^{T_g} dt  \|\vec r\|_2^2.
\end{align}
This quantity provides supplementary information to the \textit{filtering order} \cite{PAZ-SILVAGeneralTransferFunctionApproachNoise2014}, which captures the low-frequency behavior of the filter function. 
In a manner analogous to the filter flatness condition for time-dependent noise expressed in the framework of SCQC~\cite{LIDesigningArbitrarySingleaxisRotations2021}, the CFI quantifies the ability of the curve to filter low-frequency dephasing noise, provided that the bulk of it is suppressed, a condition that is achieved when the curve is closed. 
Intuitively, a smaller CFI should indicate filter functions with weight at high frequencies, since the CFI is proportional to the overlap of the filter function with the $\omega^{-2}$ PSD. In Fig.~\ref{td_ff_plot}a, we validate this expectation by plotting the filter functions for both designed gates using the \texttt{filter\_functions} package \cite{HANGLEITERFilter_functions2024}. 
It is evident from Fig.~\ref{td_ff_plot}a that the Hadamard's filter function has more weight at high frequencies, while Fig.~\ref{td_ff_plot}b (the inset) shows it also has less overlap with a unit-strength $\omega^{-2}$ PSD. This behavior is consistent with the 1.5$\times$ smaller CFI for the Hadamard compared to the $X$ gate, as obtained from Eq.~\eqref{eq:cfi}.

We conclude this section by simulating the derived controls under time-dependent noise generated with the method described in Appendix~\ref{noise_gen_calc} for $\alpha=1,2$  and compare the performance of the BARQ-optimized gates. Given the behavior of the filter functions, we expect the Hadamard gate to be less sensitive to noise for both values of $\alpha$.
In Fig.~\ref{td_ff_plot}c, we show the gate infidelity for several noise strengths expressed in terms of the dimensionless quantity $T_g\delta_z$. The infidelities are averaged over 10,000 noise realizations.

The general trend is that the analytic predictions follow the simulations closely, with the Hadamard appearing less sensitive to noise, as predicted from the filter functions in Fig~\ref{td_ff_plot}a. Discrepancies from the expected behavior can be attributed to the number of noise realizations used in the simulations and potential errors introduced from the filtering process employed for the generation of the noise samples. For increasing noise strength, larger deviations are natural since the results are correct only to first order, under the assumption that $T_g\delta_z$ is sufficiently small, a condition which becomes less valid as the noise strength increases in Fig.~\ref{td_ff_plot}c. This phenomenon has the strongest effect on the performance of the $X$ gate for $\alpha=2$. Another potential source of prediction error is the calculation of the overlap integral in Eq.~\eqref{infid_dephasing}, whose integrand diverges at zero frequency, which can lead to numerical inaccuracies in the result. We additionally note that the absolute error between the calculation of the infidelity using the CFI compared to the calculation provided from the \texttt{filter\_functions} package is on the order of $10^{-4}$.

\section{Conclusion}
\label{conclusion}

How we choose to describe or parameterize quantum evolution can significantly influence how effective a pulse design procedure is or how challenging it is to implement. Efficacious choices can avoid unnecessary tradeoffs between target objectives or save time by narrowing an otherwise prohibitively large parameter space to search through. In this work, we leveraged the SCQC formalism to devise a systematic method to design experimentally friendly and noise-suppressing quantum control pulses using geometry. We showed that by recasting quantum evolution into space curves and converting the design of dynamically corrected gates into a curve construction problem, we can automatically achieve many of the target objectives by fixing certain local properties of the space curve upfront. These include ensuring that the final evolution exactly equals the target gate in the absence of noise, suppressing first-order dephasing noise, and enforcing some pulse shape constraints. This is unlike most numerical approaches to designing control fields, where such properties are instead included as soft constraints that are imposed only at the level of a cost function. A key feature of our approach is the use of the adjoint representation, which has not previously been considered in works on SCQC. 

By combining our version of SCQC together with the concept of Bézier curves and a trick we introduced called total torsion compensation, we formulated a general numerical procedure called BARQ that automates the design of space curves that yield control fields which generate noise-robust target gates. We showed that the use of Bézier curves has the advantage of allowing us to build-in many of the desired properties simply by setting a few control points appropriately. A small number of additional control points are then optimized to achieve remaining objectives. In this way, we significantly shrink the optimization space and eliminate any tradeoffs between gate-fixing and noise-robustness. Because BARQ utilizes control points as the central element for pulse design, it can be regarded as a hybrid method in which smooth continuous pulses are generated via a discrete optimization procedure.

We demonstrated the BARQ design flow by applying it to find pulses that exhibit particular noise-robustness properties, possess experimentally-friendly characteristics, and which generate robust $X$ and Hadamard gates. Our simulation results demonstrated that BARQ can easily find pulses that simultaneously suppress two types of quasi-static noise, and we assessed the performance of these BARQ-optimized pulses against low-frequency dephasing noise. Our proposed method, along with the SCQC formalism, are implemented in the software package \texttt{qurveros}, available at \url{https://github.com/evpiliouras/qurveros}.

Although our work focused on single-qubit robust quantum control design, BARQ can be naturally extended to higher dimensions, as was already done for SCQC in Ref.~\cite{BUTERAKOSGeometricalFormalismDynamicallyCorrected2021}. Using the transfer matrix formalism as a guide to connect the dynamics of a higher-dimensional Frenet-Serret frame with the quantum evolution, the higher-dimensional control points of the associated Bézier curve will play a similar role fixing the target gate and achieving noise-robustness. In regards to the optimization step of BARQ, there is more space for developments, including but not limited to pulse characteristics. For instance, by studying how the choice of the PGF and the PRS can further simplify the optimization stage and perhaps encode pulse features through discrete symmetries, a more informed ansatz can be constructed in a manner that is analogous to guaranteeing robustness against dephasing noise by enforcing the closed-curve constraint. In this work, we limited our choices to simple examples in order to demonstrate the ability of BARQ to find the desired pulses without sophisticated constraints in the optimization phase.

\section*{ACKNOWLEDGMENTS}

This work was supported by the Department of Energy  (Grant No. DE-SC0022389). EB also acknowledges support from the Office of Naval Research (Grant No. N00014-25-1-2125).

\appendix
\section{Derivation of the equations involved in the SCQC formalism}
\label{scqc_app_equations}

In this section, we derive the equations that relate quantum evolution to geometric space curves. In SCQC, we define the space curve in terms of the error in the evolution operator caused by a Hamiltonian noise term $H_\mathrm{n}$. In this work, we focus on single-qubit evolution subject to dephasing noise with $H_n\sim\sigma_z$, and we define the space curve's tangent vector $\vec{T}$ to be the integrand of the leading-order noise error~\cite{ZENGGeometricFormalismConstructingArbitrary2019}:
\begin{align}
    \vec{T} \cdot \sigvec = U_0^\dagger \, (\hat{z} \cdot \sigvec)\, U_0,
\end{align}
where $\sigvec = \begin{bmatrix}
    \sigma_x &\sigma_y&\sigma_z\end{bmatrix}^T$. There are other ways to define space curves~\cite{YIRobustQuantumGatesCorrelated2024}, but we find the present definition most useful for designing DCGs. 
We define $\sigvec_{U_0} = U_0^\dagger \,\sigvec \,U_0$, so that we can compactly write:
\begin{align}
    \vec{T} \cdot \sigvec = \hat{z}\cdot\sigvec_{U_0}.
    \label{app1: tangent}
\end{align}
Using properties of the Pauli algebra, especially the fact that  $[\vec a \cdot \sigvec, \vec b \cdot \sigvec] = 2i (\vec a \times \vec b) \cdot \sigvec$, we find that for a constant vector $\vec a$, we have:
\begin{align}
    \frac{d}{dt} \vec a  \cdot  \sigvec_{U_0} = -i U_0^\dagger[\vec a\cdot \sigvec,H_0] U_0 = (\vec a\times \vec h_0)  \cdot  \sigvec_{U_0},
\end{align}
where $H_0 = \frac{1}{2} \vec h_0 \cdot \sigvec$.
This property allows us to interpret the time derivative as a rotation in the parameter space of the control. The interpretation of dephasing robustness as a closed curve condition, for every curve parameterization, leads to regarding time as the arclength parameterization of the curve, formally expressed as:
\begin{align}
    \frac{dt}{dx} = \left\|\frac{d\vec{r}}{dx}(x) \right\|_2 = \gamma(x),
\end{align}
where $\vec r(x)$ is a space curve, and $x$ is an arbitrary parameterization of the curve. From this condition, time is equivalent to the distance traveled along the curve $\vec r$. The \textit{normalized} tangent vector is then defined as:
\begin{align}
    \vec T = \frac{\frac{d\vec{r}}{dx}(x)}{\gamma(x)}
    \label{app1:tangent_def}.
\end{align}
All curves used in SCQC are assumed to be regular, a condition which requires the speed to remain nonzero throughout the traversal \cite{SHIFRINDIFFERENTIALGEOMETRYFirstCourse2024}.

To continue the analysis, we differentiate Eq.~\eqref{app1: tangent} twice with respect to the time variable:
\begin{align}
    \dot{\vec T} \cdot \sigvec &= (\hat z \times \vec h_0)\cdot \sigvec_{U_0}  = \Omega \,\hat \Phi \cdot \sigvec_{U_0},
    \label{tdot_eq}\\
    \ddot{\vec T} \cdot \sigvec &= (\dot \Omega \hat \Phi-\Omega\dot\Phi\hat\rho_\Phi + \Omega \,\hat{\Phi}\times \vec h_0) \cdot \sigvec_{U_0} \nonumber\\
    &=(-\Omega^2\hat{z} -\Omega(\dot\Phi -\Delta)\hat\rho_\Phi +\dot\Omega \hat \Phi)\cdot\sigvec_{U_0}
    \label{tdoubledot_eq}
\end{align}
where the Hamiltonian was expressed in cylindrical coordinates as $\vec h_0 = \Omega \,\hat {\rho}_\Phi + \Delta \hat z$ with $\hat \Phi = -\sin\Phi \hat x + \cos\Phi \hat y$. We can additionally take the commutator between Eq. \eqref{app1: tangent} and  Eq. \eqref{tdot_eq} to obtain:
\begin{align}
    (\vec T \times \dot {\vec{T}})\cdot\sigvec = -\Omega  \hat\rho_\Phi \cdot\sigvec_{U_0}
    \label{tcrosstdot_eq}
\end{align}

To find the relations between control fields and the curvature and torsion of the space curve, we multiply Eq.~\eqref{tdot_eq} separately with itself and with  Eq.~\eqref{tcrosstdot_eq}. Taking the trace, we find
\begin{align}
    \Omega^2 &= \|\dot {\vec{T}}\|^2_2, \\
\Omega^2 (\dot\Phi-\Delta) &= (\vec T \times  \dot{\vec T})\cdot \ddot{\vec T}.
\end{align}
When $\|\dot {\vec{T}}\|\neq 0$, the equations of the main text are recovered.
We observe that the envelope of the driving field is not enforced.
In the following section, we will discuss the choice of the sign and analyze points where the mapping is not well-defined.

Note that the dephasing robustness condition is recovered by integrating Eq.~\eqref{app1: tangent}. The closed curve condition is then expressed as $\vec r(T_g)=\vec r(0)$. In the main text, we assumed $\vec r(0)=\vec 0$ without loss of generality.

\section{Inflection points and parameterization-independent frame calculations}
\label{scqc_app_mvframe_calc}

\subsection{The continuous Frenet-Serret frame}
In the previous section, we derived equations that relate control fields to derivatives of a space curve. To further sharpen this relationship, we first need to address some subtleties related to the Frenet-Serret (FS) frame. Normally, this frame is defined such that the curvature of the space curve is given by $\|\dot {\vec T}\|_2$, which is strictly nonnegative, while the normal vector is $\dot {\vec T}/\|\dot {\vec T}\|_2$. However, the latter can have jump discontinuities at points where $\dot {\vec T}=0$, i.e., \emph{inflection points}, at which the curvature also vanishes. This can be seen by employing the concept of \textit{lines} as studied in Ref.~\cite{HORDTorsionInflectionPointSpace1972};  using a Taylor expansion near an inflection point $t_s$ we find:
\begin{align}
    \dot {\vec T}/\|\dot {\vec T} \|_2= (\text{sgn}(t-t_s))^{l_s} \frac{\dot{\vec{T}}^{(l_s)}}{\|\dot{\vec{T}}^{(l_s)}\|_2},\qquad t \neq t_s,
    \label{tdot_normed_inflection}
\end{align}
where $l_s$ is the order of the lowest-order nonzero derivative at inflection point $t_s$, and $\text{sgn}(\cdot)$ is the sign function. Note that at an inflection point, higher-order derivatives of the tangent vector could also vanish: $\dot{\vec{T}},\dot{\vec{T}}^{(1)},\dots,\dot{\vec{T}}^{(l_s - 1)} = 0,\,l_s >0$. If the order $l_s$ of the lowest nonzero derivative is odd, $\dot {\vec T}/\|\dot {\vec T}\|_2$ is discontinuous. For the purposes of this work, we call such a point \textit{singular}.

However, the presence of discontinuities at singular points would be at odds with Eqs.~\eqref{preadj_B}--\eqref{preadj_T}, from which it can be seen that if the evolution operator and the phase field $\Phi(t)$ are continuous, then the FS frame vectors should also be continuous. To reconcile this issue, we abandon the conventional definition of the FS frame and instead allow the curvature to assume both positive and negative values so that the FS frame is continuous even at inflection points. For these reasons, in this work we define the curvature as
\begin{align}
    \kappa(t) = f(t) \|\dot {\vec T}(t)\|_2,
\end{align}
where $f(t)$ is a switching function that changes sign after a singular point, defined as
\begin{align}
    f(t) = \sum_{m=0}^{M} (-1)^{m}(\Theta(t-t_m) - \Theta(t-t_{m+1})).
\end{align}
Here, $M$ is the number of singular points on the curve, $\Theta (\cdot)$ is the step function, and $t_0=0$ and $t_{M+1}=T_g$.
Note that we have defined $f(t)$ such that it has the initial value $f(0)=1$; this means that the continuous FS frame at $t=0$ coincides with the conventional FS frame. Using this signed curvature and continuous FS frame allows us to directly equate the envelope field to the curvature:
\begin{align}
    \Omega(t)=\kappa(t).
\end{align}
Note that a singular point can be detected by either finding the order of the lowest-order nonzero derivative at an inflection point or by simply tracking numerically the direction of $\dot T(t)$ before and after a singular point, since based on Eq.~\eqref{tdot_normed_inflection}, the sign change is reflected in the direction of the vector $\dot T(t)$ \cite{HORDTorsionInflectionPointSpace1972}.

We can now define the continuous normal and binormal FS frame vectors as
\begin{align}
    \vec{N}(t) &= \dot {\vec T}(t)/\kappa(t),\\
    \vec{B}(t) &= \vec T(t) \times \vec{N}(t).
\end{align}
To determine a frame vector at an inflection point, we can take the limit from either above or below that point, unless the point in question is an endpoint. For an inflection point at $t=0$, we have:
\begin{align}
    \vec N(0)=\lim_{t\to0^+} \frac{1}{f(t)}  \frac{\dot {\vec{T}}}{\|\dot {\vec{T}}\|_2}= \left.\frac{\dot{\vec{T}}^{(l_s)}}{\|\dot{\vec{T}}^{(l_s)}\|_2}\right|_{t=0},
\end{align}
and so the normal vector is continuous at $t=0$ regardless of $l_s$. If there is an inflection point at $t=T_g$, we find:
\begin{align}
    \vec N(T_g)=\lim_{t\to T_g^-}  \frac{1}{f(t)}  \frac{\dot {\vec{T}}}{\|\dot {\vec{T}}\|_2}=\frac{(-1)^{l_s}}{f(T_g^-)}\left . \frac{\dot{\vec{T}}^{(l_s)}}{\|\dot{\vec{T}}^{(l_s)}\|_2}\right|_{t=T_g}.
\end{align}
We notice that the sign factor depends on the total number of singular points $M$. This is a general behavior since the value of $f(t_s^-)$ contains all the sign changes induced up to the inflection point in question.

With these definitions, we can write Eq.~\eqref{tdot_eq} and Eq.\eqref{tcrosstdot_eq} as
\begin{align}
    \vec N \cdot \sigvec &= \hat \Phi \cdot \sigvec_{U_0}\label{cont_normal_eq}, \\
    -\vec B \cdot \sigvec &= \hat \rho_\Phi \cdot \sigvec_{U_0}.
    \label{cont_binormal_eq}
\end{align}
To find the second field condition, we differentiate Eq.~\eqref{cont_normal_eq}:
\begin{align}
    \dot{\vec N} \cdot \sigvec = -(\Omega\hat z+(\dot\Phi-\Delta)\hat\rho_\Phi)\cdot \sigvec_{U_0}.\label{cont_normal_dot}
\end{align}
The field condition is recovered by multiplying Eq.~\eqref{cont_binormal_eq} with Eq.~\eqref{cont_normal_dot} to find:
\begin{align}
    \dot \Phi - \Delta = \dot {\vec N}\cdot\vec B = \tau.
\end{align}
where $\tau$ is the torsion. 

In Ref.~\cite{HORDTorsionInflectionPointSpace1972}, it is shown that the torsion function is an analytic function with respect to the arclength, even in the presence of inflection points. The definition of the continuous FS frame ensures the smoothness of $ \dot {\vec N}$. We differentiate Eq.~\ref{cont_normal_eq} to obtain
\begin{align}
    \dot {\vec N} = \frac{\kappa\ddot{\vec T}-\dot\kappa\dot{\vec T}}{\kappa^2}=\frac{\ddot{\vec T}-(\ddot{\vec T}\cdot \vec N)\vec N}{\kappa}.
\end{align}
When the torsion function is evaluated at a regular point, we find the standard expression as
\begin{align}
    \tau(t) = \frac{\vec T \times \dot{\vec T}}{\|\vec T\times \dot {\vec T}\|_2^2}\cdot \ddot{\vec T}
\end{align}
At an inflection point $t_s$, we can use the limit as $t\to t_s$ to calculate the torsion as
\begin{align}
    \tau(t_s) &= \lim_{t\to t_s} \dot {\vec N}(t)\cdot\vec B(t) = \lim_{t\to t_s} \frac{\vec B}{\kappa}\cdot \ddot{\vec T}\nonumber\\
    &=\lim_{t\to t_s} \frac{\vec T \times \dot {\vec T}}{\kappa^2}\cdot\ddot{\vec T}\nonumber\\
    &=\frac{\vec T \times \dot{\vec{T}}^{(l_s)}}{\|\vec T \times \dot{\vec{T}}^{(l_s)}\|_2^2}\cdot\frac{\dot{\vec{T}}^{(l_s+1)}}{l_s + 1}.
\end{align}
which follows from Taylor-expanding $\vec B$ and $\ddot{\vec T}$ about $t=t_s$ and noticing that $(\vec T \times \dot{\vec{T}}^{(l_s)})\cdot\dot{\vec{T}}^{(l_s)} =0$. For $l_s = 0$, the standard expression for torsion is recovered. A derivation using L'Hopital's rule is found in \cite{HORDTorsionInflectionPointSpace1972}. 

When calculations are made in the neighborhood of a singular point, the sign change of the envelope equivalently imparts an additional order-dependent phase. For convenience, we can define an effective phase field from the term $f(t)\hat\Phi$ in Eq.~\eqref{cont_normal_eq} as:
\begin{align}
    \Phi_{\text{eff}} =  \Phi + \pi \sum_{m=1}^{M} \Theta(t-t_m).
    \label{effective_phase_field}
\end{align}
The definition of the effective phase field enters similarly with the gauge transformation discussed in Refs.~\cite{HUDiscreteFrenetFrameInflection2011, SHIFRINDIFFERENTIALGEOMETRYFirstCourse2024, BISHOPThereMoreOneWay1975}.

If the Taylor expansion method in Ref.~\cite{HORDTorsionInflectionPointSpace1972} is not employed, we can alternatively use Eq. \eqref{tdot_eq} to find the FS frame at an inflection point. We can proceed by differentiating Eq. \eqref{tdot_eq} as many times as necessary until the vector quantity $\vec T \times \dot{\vec{T}}^{(l_s)}$ does not vanish.
\subsection{Calculation of the frame equations in non-arclength parameterization}
In this section, we discuss an alternative route to calculate the FS frame and the associated geometric quantities. While the analysis involving the inflection points was made using the normal vector, we will instead focus on the binormal vector, since the cross product simplifies some calculations. At any point, be it regular or an inflection point, we will focus on the quantity
\begin{align}
    \frac{\vec T \times \dot{\vec{T}}^{(l_s)}}{\|\vec T \times \dot{\vec{T}}^{(l_s)}\|_2},
\end{align}
where we assume that $\dot{\vec{T}}^{(l_s)} \neq \vec 0$ and derivatives of lower order vanish. We write Eq. \eqref{app1:tangent_def} in the form $\frac{d\vec r}{dx} = \gamma(x) \vec T$, and differentiate up to the order $m$. We find a general expression of the form
\begin{align}
 \frac{d^m\vec r}{dx^m} =
  \frac{d^{m-1}\gamma}{dx^{m-1}}
 \vec T + \gamma^m \dot{\vec T}^{(m-2)} + \mathcal{O}(\dot{\vec T}^{(m-3)}),
\end{align}
for $m\ge3$.
Assuming that the highest nonzero derivative of $\dot{\vec{T}}$ is $l_s$ at an inflection point $t=t_s$, we find
\begin{align}
    \frac{d^{l_s+2}\vec r}{dx^{l_s+2}} =
  \frac{d^{l_s+1}\gamma}{dx^{l_s+1}}
 \vec T + \gamma^{l_s+2} \,\dot{\vec T}^{(l_s)}, \\
   \frac{d^{l_s+3}\vec r}{dx^{l_s+3}} =
  \frac{d^{l_s+2}\gamma}{dx^{l_s+2}}
 \vec T + \gamma^{l_s+3} \,\dot{\vec T}^{(l_s+1)}+\mathcal{O}(\dot{\vec T}^{(l_s)}). 
\end{align}
When the differentiation variable is not explicitly defined, the derivative is assumed to be with respect to time.
We observe that when an inflection point is met, the derivatives of the space curve become parallel to the tangent, up to a specific order. At an inflection point, we can equivalently use the following equations, which are not expressed in the time variable:
\begin{align}
    \frac{\vec T \times \dot{\vec{T}}^{(l_s)}}{\|\vec T \times \dot{\vec{T}}^{(l_s)}\|_2} &= \frac{\frac{d\vec r}{dx} \times\frac{d^{l_s+2}\vec r}{dx^{l_s+2}} }
    {\|\frac{d\vec r}{dx} \times\frac{d^{l_s+2}\vec r}{dx^{l_s+2}}\|_2}, \label{nonarc_unsigned_binormal} \\
   \frac{\vec T \times \dot{\vec{T}}^{(l_s)}}{\|\vec T \times \dot{\vec{T}}^{(l_s)}\|_2^2}\cdot\dot{\vec{T}}^{(l_s+1)} &= \frac{\frac{d\vec r}{dx} \times\frac{d^{l_s+2}\vec r}{dx^{l_s+2}} }
    {\|\frac{d\vec r}{dx} \times\frac{d^{l_s+2}\vec r}{dx^{l_s+2}}\|_2^2}\cdot \frac{d^{l_s+3}\vec r}{dx^{l_s+3}} .\label{nonarc_torsion} 
\end{align}
The above expressions are valid even for $l_s =0$, which does not correspond to an inflection point.

From an algorithmic point of view, one needs to evaluate sequentially the quantity $\|r'\times r{''}^{(q)}\|_2, \, q\ge0$, where the prime denotes the derivative with respect to the non-arclength parameter. When a $q^*=l_s$ is found such that the defined quantity does not vanish, we then evaluate the quantities in Eqs.~\eqref{nonarc_unsigned_binormal} and 
\eqref{nonarc_torsion}. We can then complete the calculation of the frame using the fact that $\vec B = \vec T \times \vec N$ and $\vec T \cdot \dot{\vec{T}}^{(l_s)}=0$ at an inflection point.
Special care must be taken for the numerical accuracy of the algorithm that detects inflection points, since the quantity $\|r'\times r'^{(q)}\|_2$ might cause floating point errors.
\section{SCQC under the Liouville representation}
\label{liouville_scqc}
\subsection{ 
Adjoint representation and the Liouville representation}
In this section, we provide the calculations that connect the adjoint representation of SCQC with the Liouville representation or transfer matrix representation 
\cite{CHOWUniversalQuantumGateSet2012,
GILCHRISTVectorizationQuantumOperationsIts2009,
GREENBAUMIntroductionQuantumGateSet2015,
HANGLEITERFilterfunctionFormalismSoftwarePackage2021,
NIELSENGateSetTomography2021,
HASHIMPracticalIntroductionBenchmarkingCharacterization2024}. This concept is also known as the superoperator formalism. The main idea is that, given a $d$-dimensional quantum system and a quantum state $\rho$, we can find a representation of the action of a quantum channel $\Lambda(\rho)$ on the state $\rho$. In the context of quantum tomography, it was found that this approach significantly simplifies the formulation of the problem \cite{GREENBAUMIntroductionQuantumGateSet2015, NIELSENGateSetTomography2021}.

A $d$-dimensional system is equipped with $d^2$ operators $C_i$ that form a complete set and can be chosen orthonormal. We can fully characterize the action of the channel by studying its effect on every operator. Analogously to vector transformations, we shall calculate the equivalent of projections, where, in this context, they are expressed using the Hilbert-Schmidt product for two operators $A,B$, defined as $\text{tr}(A^\dagger B)$. We will define the transfer matrix $S_\Lambda$, as the matrix that contains the aforementioned information:
\begin{align}
    S_\Lambda^{ij} = \text{tr}(C_i^\dagger \Lambda(C_j)).
\end{align}
When the operators $C_i$ are chosen as the Pauli operators, we refer to it as the Pauli Transfer Matrix (PTM). 

To provide a sense of matrix operations, we associate the operators $C_i$ with their \textit{superket} counterpart $\ket{C_i}\rangle \equiv\ket{i}\rangle$. We refer the reader to this excellent review \cite{GYAMFIFundamentalsQuantumMechanicsLiouville2020} for a more detailed approach to the matter. The utility of this method arises from the fact that the Hilbert-Schimdt inner products are converted to the usual state overlaps between the superkets. The transfer matrix elements are equivalently written as
\begin{align}
     S_\Lambda^{ij} = \superbra{i} \Lambda(C_j) \rangle\rangle.
\end{align}
The density operator can be in principle expressed in a chosen operator basis as:
\begin{align}
    \rho = \sum_i \text{tr}(C_i^\dagger\rho)C_i.
\end{align}
In this regard, writing an operator to its superket form is a linear process, so the density operator can be expressed as
\begin{align}
    \ket{\rho}\rangle &= \sum_i \text{tr}(C_i^\dagger\rho) \superket{i} = \sum_i \superbra{i}\rho\rangle\rangle \superket{i} \nonumber\\
    &=\sum_i \superket{i}\superbra{i} \, \superket{\rho},
\end{align}
which essentially reveals the resolution of identity in the new space. Using that observation, we can describe the action of a quantum channel $\Lambda$ on the state $\rho$ as:
\begin{align}
\ket{\Lambda(\rho)}\rangle &= \sum_i \superket{i}\langle\bra{i}\Lambda(\rho)\rangle\rangle = \sum_i \langle\bra{i}\Lambda(\rho)\rangle\rangle \ket{i}\rangle \nonumber\\
&=\sum_i \text{tr}(C_i^\dagger \Lambda(\rho)) \superket{i}\nonumber\\
&=\sum_{i,j} \text{tr}(C_i^\dagger \Lambda(C_j)) \superket{i} \superbra{j}\rho\rangle\rangle \nonumber\\
&= \sum_{i,j} S_\Lambda^{ij} \ket{i}\rangle \langle\bra{j}\rho\rangle\rangle \nonumber\\
&=S_\Lambda \superket{\rho}.
\end{align}
In case that the quantum channel acts in a manner similar to evolving a quantum state with unitary dynamics, the transfer matrix can be regarded as a \textit{superpropagator} $\mathcal{U}$, with elements given by:
\begin{align}
    \mathcal{U}^{ij} = \langle\bra{i}UC_jU^\dagger\rangle\rangle = \text{tr}(C_i^\dagger UC_jU^\dagger)  .
\end{align}
By definition, every element for $i,j >0$ coincides with the adjoint representation of the propagator $U$, when the $C_i$ are chosen to be the Pauli operators. If we further assume physical and unital dynamics, we can write:
\begin{align}
    \mathcal{U} = \begin{bmatrix}
        1 & \hphantom{-}{0}_{d^2-1} \\
        \hphantom{-}{0}_{d^2-1} & R_U
    \end{bmatrix}.
\end{align}
While the name was motivated under the unitarity assumption, the superpropagator can capture incoherent processes, if it's regarded as the formal solution of the dynamics in the superoperator formalism.

In this work, we employed the average gate fidelity to assess the performance of the derived controls. Based on \cite{PEDERSENFidelityQuantumOperations2007}, for any complex matrix $M$, we have:
\begin{align}
   \mathcal{F}(M,I_d) = \mathcal{F}(M) = \frac{\text{tr}(MM^\dagger)+|\text{tr}(M))|^2}{d(d+1)} ,
\end{align}
where $M = U_0(T_g)U_g^\dagger$. In the transfer matrix formalism, we can perform the following calculations:
\begin{align}
  &|\text{tr}(M))|^2 = \sum_{i,j} \bra{j}M\ket{j}\bra{i}M^\dagger\ket{i}  \nonumber \\
  &=  \sum_{i,j} \text{tr}(\ket{i}\bra{j} M\ket{j}\bra{i}M^\dagger) 
  =
  \sum_{i,j} \text{tr}(\ket{i}\bra{j} \Lambda_M (\ket{j}\bra{i}))\nonumber\\
  &=\sum_{l=0}^{d^2-1}\sum_{i,j}
  \superbra{\Pi_{ji}} \Lambda_M(C_l)\rangle\rangle \superbra{l} \Pi_{ji}\rangle\rangle \nonumber\\ 
  &=\sum_{l=0}^{d^2-1}\sum_{i,j} \superbra{l} \Pi_{ji}\rangle\rangle \superbra{\Pi_{ji}} \Lambda_M(C_l)\rangle\rangle \nonumber\\ 
  &= \sum_{l=0}^{d^2-1} \superbra{l}\Lambda_M(C_l)\rangle\rangle =
  \text{tr}(\mathcal{M}),
\end{align}
where the last trace operation refers to the $d^2$-dimensional space and $\Pi_{ij} = \ket{i}\bra{j}$. The matrix $\mathcal{M}$ denotes the transfer matrix representation of $M$. The last step is proven using the resolution of identity for that space.
We have:
\begin{align}
    \rho &= \sum_{ij} \bra{i}\rho\ket{j} \ket{i}\bra{j} = \sum_{ij} \text{tr}(\ket{j}\bra{i}\rho)\ket{i}\bra{j} \nonumber\\&=\sum_{ij}\superbra{\Pi_{ij}}\rho\rangle\rangle \Pi_{ij},
\end{align}
and:
\begin{align}
    \ket{\rho}\rangle =  \sum_{ij} \superbra{\Pi_{ij}}\rho\rangle\rangle \superket{\Pi_{ij}}=
     \sum_{ij} \superket{\Pi_{ij}}  \superbra{\Pi_{ij}}\superket{\rho}.
\end{align}
The resolution of identity is expected since the $\Pi_{ij}$ can express all density operators in the $d$-dimensional space. Treating the first element with the appropriate scale of the identity, and identifying the zero superket $\ket{0}\rangle$ as the Liouville representation of the identity operator $I_d/\sqrt{d}$, we recover the expression:
\begin{align}
    {\mathcal{F}}(M) = \frac{d\mathcal{M}_{00} +\text{tr}(\mathcal{M})}{d(d+1)}.
\end{align}

When the ideal unitary dynamics $U_0$ are described through the adjoint representation $R_{U_0}$ and the target gate $U_g$ is expressed as $R_g$, the average gate fidelity is written as:
\begin{align}
    \mathcal{F}_g(U_0(T_g)) = \frac{d+1+\text{tr}(R_g^T R_{U_0}(T_g))}{d(d+1)},
\end{align}
since the dagger operation translates to a transpose operation for the superpropagator. Essentially, we can achieve unit fidelity by setting the respective adjoint representations equal.

\subsection{Dynamics of the Frenet-Serret frame in the Liouville representation}

As we have established in the previous section, the adjoint representation is a sub-block of the superpropagator. In this section, we will show that the single-qubit density operator evolves in a manner similar to the moving frame. Starting with the Liouville-von Neumann equation:
\begin{align}
    i\dot \rho = [H_0,\rho],
\end{align}
and expanding the density operator, we find:
\begin{align}
    \superket{\dot \rho} = -i \superket{[H_0,\rho]} = -i  \sum_l \superket{[H_0,C_l]}\superbra{l}\rho\rangle\rangle.
\end{align}
The generator of the dynamics, called Liouvillian \cite{GYAMFIFundamentalsQuantumMechanicsLiouville2020}, is expressed as:
\begin{align}
   \mathcal{L}_0 =   -i  \sum_l \superket{[H_0,C_l]}\superbra{l}.
\end{align}
We can identify $\mathcal{L}_0$ as the generator of the superpropagator $\mathcal{U}_0$, which is understood under the action $\superket{\rho(t)} = \mathcal{U}_0(t)\superket{\rho(0)}$. An element of the Liouvillian is given by:
\begin{align}
    \mathcal{L}_0^{kl} = -i \superbra{k} [H_0,C_l]\rangle\rangle.
\end{align}
We will choose a Hermitian orthonormal operator basis, and write the matrix elements as:
\begin{align}
  \mathcal{L}_0^{kl} &=  -i \superbra{k} [H_0,C_l]\rangle\rangle \nonumber \\ 
  &=-i \text{tr}(C_k^\dagger [H_0,C_l]) \nonumber\\
  &=-i \text{tr}(C_k H_0 C_l - C_kC_lH_0 ) \nonumber\\
  &= -i  \text{tr} ([C_l,C_k]H_0) \\
  &= -i \superbra{[C_k,C_l]} H_0\rangle\rangle
\end{align}
The Hamiltonian enters as a vector with components given by its expansion on the chosen operator basis. 

We can now calculate the Liouvillian for the single-qubit control. We have:
\begin{align}
    \superket{H_0} = \frac{1}{\sqrt{2}} \begin{bmatrix}
        0 & \Omega\cos\Phi &\Omega\sin\Phi &\Delta
    \end{bmatrix}^T.
\end{align}
For the Pauli basis, we use the Levi-Civita symbol to write:
\begin{align}
    \mathcal{L}_0^{kl} &= -i \sum_m \superbra{[C_k,C_l]} m\rangle\rangle\superbra{m} H_0\rangle\rangle \nonumber \\
    &=-\sqrt{2} \sum_m  \epsilon_{klm} \superbra{m} H_0\rangle\rangle 
\end{align}
For convenience, we follow the convention that for any zero index, the Levi-Civita symbol vanishes, since the basis elements (except for the identity) are traceless. We can find the Liouvillian as:
\begin{align}
    \mathcal{L}_0 = \begin{bmatrix}
        0 & 0 & 0 & 0 \\
        0 & 0 & -\Delta & \hphantom{-}\Omega\sin\Phi\\
        0 & \Delta & 0 & -\Omega\cos\Phi \\
        0 & -\Omega\sin\Phi & \Omega\cos\Phi & 0 \\
    \end{bmatrix}.
\end{align}
 As far as the moving frame is concerned, we rearrange the moving frame vectors and rotate the normal vectors by an angle $\eta(t)$ with respect to the tangent vector such that \cite{HUDiscreteFrenetFrameInflection2011}:
\begin{align}
    \begin{bmatrix}
        \vec e_1 \\
        \vec e_2
    \end{bmatrix} =   \begin{bmatrix}
        \cos\eta & -\sin\eta \\
        \sin\eta & \cos\eta
    \end{bmatrix}\begin{bmatrix}
        \vec N \\
        \vec B
    \end{bmatrix}.
\end{align}
The general set of equations that describe the evolution of a moving frame becomes:
\begin{align}
  \frac{d}{dt}  \begin{bmatrix}
        \vec e_1 \\
        \vec e_2 \\
        \vec T
    \end{bmatrix} =   \begin{bmatrix}
        0 & \tau - \dot \eta &-\kappa\cos\eta\\
        -(\tau - \dot\eta) & 0& -\kappa\sin\eta \\
        \kappa\cos\eta & \kappa\sin\eta& 0
    \end{bmatrix} \begin{bmatrix}
        \vec e_1 \\
        \vec e_2 \\
        \vec T
        \end{bmatrix}.
\end{align}
We notice that if we choose:
\begin{align}
    \kappa\cos\eta = -\Omega\sin\Phi, \\
    \kappa\sin\eta = \Omega\cos\Phi , \\
    \dot \eta - \Delta = \tau,
\end{align}
the dynamics can be made identical to those of the moving frame. This local rotation enables the generality of the curve framing problem, where for $\kappa(t)\neq0$ and $\eta=0$, we recover the Frenet-Serret equations. The angle behaves as a gauge freedom, similar to the phase field in SCQC. While we recover the same field conditions, if the curvature vanishes at a point, the resulting moving frame will not necessarily coincide with the one used in SCQC \cite{HUDiscreteFrenetFrameInflection2011, BISHOPThereMoreOneWay1975, SHIFRINDIFFERENTIALGEOMETRYFirstCourse2024}.

To make the connection more explicit, at any point that the envelope does not vanish, the dynamics of the spin vector coincide with the ones of the moving frame, up to an initial rotation. If the curvature vanishes, the differential equation still produces a continuous solution, which generates a continuous superpropagator, but it does not coincide with the phase-field-rotated normal and binormal vectors, whose continuity is enforced by appropriately changing the sign of the envelope.

One key feature of SCQC is that by changing the sign of the envelope when a singular point is met, we avoid integrating the moving frame equations which allows us to decompose the quantum evolution into local and global elements.
\newpage

\section{Fixing the target gate and ensuring curve closure with Bézier curves}
\label{Bézier_curve_calc}

In this section, we derive the conditions that fix the gate in BARQ, allowing us to separate gate-fixing from noise-robustness. At a minimum, by design BARQ automatically enforces the closed curve condition (which in terms of the Bézier curve parameter $x\in[0,1]$ is $\vec r(0) = \vec r(1)$), the
vanishing of the curvature at the endpoints at $x=0$ and $x=1$, and gate-fixing up to a $Z$-rotation.

Starting from the Bézier curve ansatz, Eq.~\eqref{bezier_curve_eq}, we evaluate the curve at the endpoints up to the third order derivative:
\begin{align}
    \vec r(0) = \vec w_0, \qquad \vec r(1) = \vec w_n, \\
     \frac{1}{n}\vec{r}'(0) = \vec w_1-\vec w_0, \qquad    \frac{1}{n}\vec{r}'(1) = \vec w_n - \vec w_{n-1}, \\
      \frac{1}{n(n-1)}\vec{r}''(0) = \vec w_2 -2\vec w_1  + \vec w_0 ,\\
      \frac{1}{n(n-1)}\vec{r}''(1) = \vec w_n -2\vec w_{n-1}  + \vec w_{n-2}, \\
      \frac{1}{n(n-1)(n-2)}\vec{r}'''(0) = \vec w_3 -3(\vec w_2 -\vec w_1)  - \vec w_0 ,\\
      \frac{1}{n(n-1)(n-2)}\vec{r}'''(1) = \vec w_n -3(\vec w_{n-1} -\vec w_{n-2})  - \vec w_{n-3} .
\end{align}
We can readily satisfy the closed curve condition with $ \vec w_0 = \vec w_n = \vec 0$. (We may fix the endpoints to the origin without loss of generality.) To force the curvature to vanish at $x=0,1$, we set $\vec{r}'(x) \times \vec{r}''(x) = 0$ at $x=0,1$. This leads to the conditions
\begin{align}
     \vec w_2  \parallel  \vec w_1,\qquad \vec w_{n-2}  \parallel  \vec w_{n-1}.
\end{align}
We now proceed to fix the gate using the adjoint representation. In order to calculate the adjoint representation of the evolution, we will first need to find an expression for the Frenet-Serret frame of the Bézier curve. We can readily calculate the tangent vectors at the boundaries $x=0,1$ as:
\begin{align}
     \vec T(0) = \frac{\vec w_1}{\|\vec w_1\|_2}, \qquad \vec T(1) = -\frac{\vec w_{n-1}}{\|\vec w_{n-1}\|_2}.
\end{align}
When we impose vanishing curvature at the endpoints, they become inflection points. As discussed in Appendix \ref{scqc_app_mvframe_calc}, the curvature is redefined in order to ensure the continuity of the frame. Since both boundaries are inflection points, we will calculate each binormal vector using the appropriate limit. In particular,
\begin{align}
    \vec B(0) &= \lim_{t\to 0^+} \vec T \times \vec N = \vec T(0)\times \frac{1}{f(0^+)}\left.\frac{\ddot{\vec{T}}}{\|\ddot{\vec{T}}\|_2}\right|_{t=0} \nonumber\\
    &= \left.\frac{\vec T \times \ddot{\vec{T}}}{\|\vec T\times \ddot{\vec{T}}\|_2}\right|_{t=0} = \frac{\vec r'(0)\times \vec r'''(0)}{\|\vec r'(0)\times \vec r'''(0)\|_2}\nonumber\\
    &=
    \frac{\vec w_1 \times \vec w_{3}}{\|\vec w_1 \times \vec w_{3}\|_2},
\end{align}
since $f(0^+)=1$. For the calculation at $t=T_g$, we proceed in a similar fashion:
    \begin{align}
    \vec B(T_g) &= \lim_{t\to T_g^-} \vec T \times \vec N = \vec T(T_g)\times \frac{-1}{f(T_g^-)}\left.\frac{\ddot{\vec{T}}}{\|\ddot{\vec{T}}\|_2}\right|_{t=T_g} \nonumber\\
    &= \frac{-1}{f(T_g^-)} \left.\frac{\vec T \times \ddot{\vec{T}}}{\|\vec T\times \ddot{\vec{T}}\|_2}\right|_{t=T_g}  \nonumber\\
    &=\frac{-1}{f(T_g^-)}\frac{\vec r'(1)\times \vec r'''(1)}{\|\vec r'(1)\times \vec r'''(1)\|_2}\nonumber\\
    &= \frac{-1}{f(T_g^-)}
    \frac{\vec w_{n-1} \times \vec w_{n-3}}{\|\vec w_{n-1} \times \vec w_{n-3}\|_2}.
    \label{binormal_end_evol}
\end{align}
Since $\vec w_1, \vec w_3$ are free parameters, we will assume that $\vec w_1 \nparallel \vec w_3$. 
At the end of the evolution, we have
\begin{align}
    R_F(T_g) = \begin{bmatrix}
        -\vec B(T_g) \\
        \hphantom{-}\vec N(T_g) \\
        \hphantom{-}\vec T(T_g)
    \end{bmatrix}.
\end{align}
We want to isolate the gate fixing from the sign induced by the singular points since this is unknown during the optimization step. To that end, we notice that the matrix $R_F(T_g)$ can be written as:
\begin{align}
    R_F(T_g) = R_Z((M+1)\pi)R_B(T_g),
\end{align}
where the matrix $R_B$ contains only the control points, and $R_Z(\theta)$ is the rotation matrix
\begin{align}
    R_Z(\theta) = \begin{bmatrix}
        \cos\theta & -\sin\theta & 0 \\
        \sin\theta & \cos\theta &0 \\
        0&0&1
    \end{bmatrix}.
\end{align}
The phase $M\pi$ is induced by the number of singular points, which is unknown during the optimization, and the additional phase of $\pi$ is from the overall negative sign in Eq.~\eqref{binormal_end_evol}.
In this context, we can equivalently express the adjoint representation as
\begin{align}
    R_{U_0}(T_g) &=  R_Z(\Phi(T_g))R_F(T_g)R_F^T(0) \nonumber\\
    &= R_Z(\Phi(T_g)+(M+1)\pi)R_B(T_g)R_B^T(0).
\end{align}

From our previous calculations, to maximize the average gate fidelity, we need to equate the adjoint representation of the target gate $U_g$ denoted as $R_g$ with the adjoint representation of SCQC at $t=T_g$. We have:
\begin{align}
    R_{U_0}(T_g) &= R_g \iff \nonumber \\
    R_B(T_g) &= R_Z^T(\Phi(T_g)+(M+1)\pi)R_gR_B(0).
\end{align}
$R_B(T_g)$ is determined entirely by the final few control points; we will assume they are chosen such that
\begin{align}
    R_B(T_g) = R_Z^T(\theta_B) R_g R_B(0),
    \label{bezier_app_barq_eq}
\end{align}
where $\theta_B$ is the BARQ angle defined in the main text. With this choice, the final evolution operator is the target gate up to a $Z$-rotation. We can correct for this residual $Z$-rotation, which is deterministic and proportional to the angular distance $|\theta_B - (\Phi(T_g)+(M+1)\pi)|$. This essentially provides a landscape of possible evolutions in which the final angle $\theta_B$ can potentially influence some of the pulse characteristics. We can achieve unit fidelity by imposing the condition
\begin{align}
    \theta_B -  (\Phi(T_g)+(M+1)\pi) = 2k\pi, k \in \mathbb{Z},
\end{align}
as described in the main text.

We now treat the control points that form $R_B(0)$ as free variables and set $R_{gB}(0) = R_g R_B(0)$. The resulting matrix is orthonormal as it is a product of orthonormal matrices. Let $\vec a_i$ be the $i$-th row of that matrix. From the structure of the equations, we readily identify the last row as
\begin{align}
   \frac{\vec w_{n-1}}{\|\vec w_{n-1}\|} \ = - \vec a_3. \label{bezier_app_r3}
\end{align}
This equation fixes only the direction of $\vec w_{n-1}$, while the fact that its magnitude is unconstrained can provide more flexibility in the design. Such choices are made in the context of the PGF. In our implementation, these magnitudes are indicated by the name of the vector followed by \texttt{\_fix} and are required to be positive. 

Looking at the remaining rows of $R_{gB}(0)$, we use Eq.~\eqref{bezier_app_barq_eq} to find:
\begin{align}
    R_B(T_g)\vec a_1= \begin{bmatrix}
        \cos\theta_B & -\sin\theta_B& 0
    \end{bmatrix}^T, \\ 
     R_B(T_g)\vec a_2= \begin{bmatrix}
        \sin\theta_B &\cos\theta_B& 0
    \end{bmatrix}^T.
\end{align}
This leads to the set of equations:
\begin{align}
     \frac{\vec a_1 \times \vec a_3}{\|\vec w_{n-3}\times \vec a_3\|_2} \cdot \vec w_{n-3} = \cos\theta_B, \\ 
     \frac{ \vec a_2 \times \vec  a_3}{\|\vec w_{n-3}\times \vec a_3\|_2} \cdot \vec w_{n-3} = \sin\theta_B.
\end{align}
We can expand $\vec w_{n-3}$ in the basis formed by $\vec a_i$: 
\begin{align}
    \vec w_{n-3} = q_1 \vec a_1 + q_2\vec a_2 + q_3\vec a_3.
\end{align}
We recall one definition of the determinant as $\det(R_{gB}(0))=\vec a_1\cdot (\vec a_2\times \vec a_3)$. Since the matrix is a proper rotation, $\vec a_1\cdot (\vec a_2\times\vec a_3) = 1$. While the value of $q_3$ remains undetermined, the other parameters are fixed by:
\begin{align}
        \frac{(\vec a_1 \times \vec a_3)\cdot \vec  a_2}{\sqrt{q_1^2 + q_2^2}} q_2 = \cos\theta_B, \\ 
         \frac{(\vec a_2 \times \vec a_3)\cdot \vec a_1}{\sqrt{q_1^2 + q_2^2}} q_1 = \sin\theta_B .
\end{align}
For a more compact representation, we set:
\begin{align}
    q_1 &= \lambda^+_{n-3}\sin\theta_B, \\
    q_2 &= -\lambda^+_{n-3}\cos\theta_B,\\
    q_3 &= -\lambda_{n-3}.
\end{align}
The parameters with the superscript indicate positive quantities, while the others can take arbitrary values.

In summary, the complete set of conditions that we impose on the initial and final control points in BARQ, which collectively guarantee a closed curve with vanishing curvature (and hence envelope field) at the endpoints and fix the target gate are
\begin{align}
    \vec w_0 &= \vec 0, \\
    \vec w_1 &= \lambda_1^+ \,\hat p_1,\\
    \vec w_2 &= \lambda_2 \,\hat p_1, \\
    \vec w_3 &= \lambda_3^+\, \hat p_2 + \lambda_3 \,\hat p_1,\\
     \vec w_{n-3} = \lambda_{n-3}^+(\sin\theta_B\, \vec a_1 &-\cos\theta_B\, \vec a_2) - \lambda_{n-3}\,\vec a_3,\\
     \vec w_{n-2} &= - \lambda_{n-2} \,\vec a_3,\\
    \vec w_{n-1} &= - \lambda_{n-1}^+ \,\vec a_3,\\
    \vec w_n &= \vec 0 ,
\end{align}
where $p_1, p_2$ are the first two free points.

\section{Noise induced errors and the Frenet-Serret frame}
\label{adjoint_noise_calc}

As discussed in the main text, the adjoint representation of SCQC provides a geometric insight into the behavior of the errors in the interaction picture. We will follow the notation and procedure from \cite{HANGLEITERFilterfunctionFormalismSoftwarePackage2021}, to first develop the setting for arbitrary noise entering the system Hamiltonian and later express the key quantities in terms of the adjoint representation of SCQC. 

We assume a general noise operator entering the system as:
\begin{align}
    H_{\text{n}} = \sum_\alpha b_\alpha(t)B_\alpha(t),
\end{align}
where $b_\alpha(t)$ are classical random possibly time-dependent fluctuations, and $B_\alpha(t)$ are hermitian, deterministic, in principle time-dependent operators. Moving to the interaction picture with respect to the noise-free dynamics described by $U_0$, the operator transforms as:
\begin{align}
    \tilde  H_{\text{n}} = U_0^\dagger  H_{\text{n}} U_0.
\end{align}
To examine the influence of the noise, we are interested in evaluating the average gate fidelity in its presence. In Appendix \ref{liouville_scqc}, we expressed the average gate fidelity using the transfer matrix $\mathcal{M}$ which was formed from $M=U_g^\dagger U_0(T_g)$. In this section, we will assume that the target gate is implemented up to a global phase (hence unit fidelity in the noise-free case), therefore the sole source of infidelity is from $H_\text{n}$. By decomposing the total propagator $U$ as a product $U = U_0 U_I$, we find $M=U_I(T_g)$ and the corresponding average gate fidelity writes:
\begin{align}
    \mathcal{F}_\text{n} = \frac{d\,\mathcal{U}_I^{00}+\text{tr}(\mathcal{U}_I)}{d(d+1)} .
\end{align}
In the case where the classical fluctuations $b_\alpha(t)$ are deterministic, the average gate fidelity is the same for every control experiment. When the $b_\alpha(t)$ are stochastic processes, the average gate fidelity is estimated by calculating the expected value over noise realizations. In other words, when identical systems are controlled with the only difference being the stochastic variables $b_\alpha(t)$, the final average gate fidelity is estimated as the expected value of the various gate fidelity results.

To that end, we use the expectation operator $\mathbb{E}_b$ and calculate the quantity $\mathbb{E}_b[\mathcal{U}_I]$. It can be shown that such an averaging operation can be executed using the cumulant expansion \cite{HANGLEITERFilterfunctionFormalismSoftwarePackage2021, CERFONTAINEFilterFunctionsQuantumProcesses2021}, which leads to the result:
\begin{align}
    \mathbb{E}_b[\mathcal{U}_I] = e^{\mathcal{K}(t)},
    \label{exp_cumulant}
\end{align}
where $\mathcal{K}$ is the cumulant function. An element of the cumulant function is given by:
\begin{align}
    \mathcal{K}_{ij} = -\frac{1}{2}\sum_{\alpha,\beta} \sum_{k,l}( f_{ijkl} \Delta_{\alpha\beta,kl} + g_{ijkl} \Gamma_{\alpha\beta,kl}),
\end{align}
where, with the basis operators $C_i$ defined in Appendix \ref{liouville_scqc}, the coefficients $f_{ijkl}, g_{ijkl}$ are defined as:
\begin{align}
    f_{ijkl} &= T_{klji} - T_{lkji}-T_{klij}+T_{lkij}, \\
    g_{ijkl} &= T_{klji} - T_{kjli}-T_{kilj}+T_{kijl}, \\
    T_{ijkl} &= \text{tr}(C_iC_jC_kC_l).
\end{align}
The effect of noise is captured by the decay amplitudes $\Gamma_{\alpha\beta,kl}$ and the frequency shifts $\Delta_{\alpha\beta,kl}$, where the first two indices describe the contribution of a pair of noise sources. The elements of those fourth-order tensors are given by:
\begin{align}
    &\Gamma_{\alpha\beta,kl} = \int_0^{T_g} \int_0^{T_g}dt_1dt_2 \,C_{\alpha\beta}(t_1,t_2)\mathcal{\tilde{ B}}_{\alpha k}(t_1)\mathcal{\tilde{ B}}_{\beta l}(t_2), \\
    &\Delta_{\alpha\beta,kl} = \int_0^{T_g} dt_1\int_0^{t_1}dt_2 \,C_{\alpha\beta}(t_1,t_2)\mathcal{\tilde{ B}}_{\alpha k}(t_1)\mathcal{\tilde{ B}}_{\beta l}(t_2),
\end{align}
where $C_{\alpha\beta}(t_1,t_2) = \mathbb{E}_b[b_\alpha(t_1)b_\beta(t_2)]$
is the cross-correlation function of the classical fluctuations and $\mathcal{\tilde{ B}}_{\alpha k}$ are the expansion coefficients of the noise operators in the interaction picture. Using  the basis of $C_i$, they are defined as:
\begin{align}
    \mathcal{\tilde{ B}}_{\alpha k}(t) = \text{tr}(\tilde B_\alpha(t)C_k) = 
   \text{tr}(U_0^\dagger B_\alpha(t)U_0C_k).
\end{align}
To simplify the calculations, we will use the transfer matrix formalism from Appendix \ref{liouville_scqc}. The coefficients $\mathcal{\tilde{ B}}_{\alpha k}$ can be expressed using the superpropagator as:
\begin{align}
    \mathcal{\tilde{ B}}_{\alpha k}(t) &=  \text{tr}(B_\alpha(t)U_0C_kU_0^\dagger)\nonumber\\ &=\superbra{B_{\alpha}(t)}U_0C_kU_0^\dagger\rangle\rangle\nonumber\\
    &=\superbra{B_{\alpha}(t)}\mathcal{U}_0(t) \superket{k}.
\end{align}
where $\superket{k}$ is the Liouville representation of the operator $C_k$ with $\langle\langle l\superket{k}=\text{tr}(C_l^\dagger C_k)$. Using the resolution of identity, we can equivalently express the product $\mathcal{\tilde{ B}}_{\alpha k}(t_1)\mathcal{\tilde{ B}}_{\beta l}(t_2)$ as:
\begin{align}
    &\mathcal{\tilde{ B}}_{\alpha k}(t_1)\mathcal{\tilde{ B}}_{\beta l}(t_2) = \superbra{B_\alpha(t_1)}\mathcal{U}_0(t_1) \superket{k}\superbra{B_\beta(t_2)}\mathcal{U}_0(t_2) \superket{l} \nonumber\\
    &= \sum_{n, m} \superbra{B_\alpha(t_1)}
    n\rangle\rangle
    \mathcal{U}_0^{nk}(t_1) \superbra{B_\beta(t_2)}m\rangle\rangle\mathcal{U}_0^{ml}(t_2)
    \nonumber\\
    &= \sum_{n,m} A_{\alpha n}(t_1)\mathcal{U}_0^{nk}(t_1) A_{\beta m}(t_2)\mathcal{U}_0^{ml}(t_2) \nonumber \\
    &=(A\mathcal{U}_0)_{\alpha k}(t_1) (A\mathcal{U}_0)_{\beta l}(t_2), \label{int_pic_coeff}
\end{align}
where the matrix $A$ contains the deterministic amplitudes of the noise operators as
$A_{\alpha k} = \langle\bra{B_\alpha} k\rangle\rangle 
$. 

In order to recover an average gate fidelity estimate, the exponential in Eq. \eqref{exp_cumulant} is expanded to first order, assuming operation in the low-noise regime. The average gate fidelity becomes:
\begin{align}
    \mathcal{F}_\text{n} = \frac{d+d^2+\text{tr}(\mathcal{K})}{d(d+1)}=1 + \frac{\text{tr}(\mathcal{K})}{d(d+1)}
\end{align}
since from the model assumptions $\mathcal{U}_I^{00}=1$, and $\text{tr}(I_{d^2})=d^2$. We can show that \cite{HANGLEITERFilterfunctionFormalismSoftwarePackage2021}:
\begin{align}
    \text{tr}(\mathcal{K}) = -d \sum_{\alpha, \beta} \sum_{k} \Gamma_{\alpha\beta,kk} = -d \sum_{\alpha, \beta} \Gamma^c_{\alpha\beta}
\end{align}
where $\Gamma^c_{\alpha\beta}$ are the contracted decay amplitudes.
In this first-order approximation, there is no contribution from the frequency shifts. We can evaluate the product of the interaction picture coefficients from Eq. \eqref{int_pic_coeff} under the tensor contraction as:
\begin{align}
    &\sum_k\mathcal{\tilde{ B}}_{\alpha k}(t_1)\mathcal{\tilde{ B}}_{\beta l}(t_2)= \sum_k (A\mathcal{U}_0)_{\alpha k}(t_1) (A\mathcal{U}_0)_{\beta k}(t_2) \nonumber \\
    &=[A(t_1)\mathcal{U}_0(t_1)\mathcal{U}^T_0(t_2)A^T(t_2)]_{\alpha\beta} \nonumber, \\
    &= [A(t_1)\mathcal{U}_{F}(t_1)\mathcal{U}^T_{F}(t_2)A^T(t_2)]_{\alpha\beta} ,
\end{align}
where $\mathcal{U}_0 = \mathcal{U}_{F}\mathcal{U}^T_{F}(0)$ and $\mathcal{U}_{F}$ is the superpropagator resulting from $U_{F}$. Since the rows are orthonormal, we have $\mathcal{U}^T_{F}(0)\mathcal{U}_{F}(0) = I_{d^2}$. Intuitively, the noise cancellation properties of a curve should not depend on the initial conditions, an expectation that is validated with the above calculations. We can then define the \textit{control-amplitude} matrix $\mathcal{A}_{F}(t)=A(t)\mathcal{U}_{F}(t)$ and compactly express the contracted decay amplitudes as:
\begin{align}
    \Gamma^c_{\alpha\beta} = \int_0^{T_g} \int_0^{T_g}dt_1dt_2 \,C_{\alpha\beta}(t_1,t_2) [\mathcal{A}_{F}(t_1)\mathcal{A}_{F}^T(t_2)]_{\alpha\beta}.
\end{align}

If we further assume a wide-sense stationary (WSS) process, we can introduce the noise Power Spectral Density (PSD) and write:
\begin{align}
    C_{\alpha\beta}(t_1,t_2)=\frac{1}{2\pi}\int_{-\infty}^{\infty}d\omega S_{\alpha\beta}(\omega)e^{-i\omega(t_1-t_2)}.
\end{align}
Using the Fourier transform of the control-amplitude matrix defined as:
\begin{align}
    \mathcal{A}_{F}(\omega) = \int_0^{T_g} dt \,\mathcal{A}_{F}(t) e^{-i\omega t},
\end{align}
the contracted decay amplitudes are compactly expressed as:
\begin{align}
    \Gamma^c_{\alpha\beta} = \frac{1}{2\pi}\int_{-\infty}^{\infty}d\omega S_{\alpha\beta}(\omega)[\mathcal{A}_{F}(\omega)\mathcal{A}_{F}^T(-\omega)]_{\alpha\beta}.
\end{align}
In this notation, the fidelity filter function for a single noise source is recovered by setting $\alpha=\beta$ and is expressed as:
\begin{align}
    F_{\alpha}(\omega) = [\mathcal{A}_{F}(\omega)\mathcal{A}_{F}^T(-\omega)]_{\alpha\alpha}.
\end{align}
In that case, the average gate infidelity becomes:
\begin{align}
    \mathcal{I}_n^{\alpha} = \frac{1}{2\pi(d+1)} \int_{-\infty}^{\infty}d\omega\, S_{\alpha}(\omega) F_{\alpha}(\omega),
\end{align}
where we set $S_{\alpha\alpha} = S_\alpha$.

\section{Derivation of the curve filtering index}
\label{cfi_calculation}

We assume an error of the form $H_\text{n} = \delta_z(t)\frac{\sigma_z}{2}$ and compute the amplitude matrix as
\begin{align}
    A = \begin{bmatrix}
        0 & 0 & 0 & \frac{1}{\sqrt{2}}
    \end{bmatrix},
\end{align}
where we assumed the basis is arranged in the sequence $I,X,Y,Z$. The last row of the superpropagator $\mathcal{U}_{F}$ coincides with the last row of the matrix $R_F$, therefore the dephasing filter function will contain the tangent vector. The control-amplitude matrix becomes
\begin{align}
    \mathcal{A}_{F}^T(t)= \frac{\vec T}{\sqrt{2}} .
\end{align}
By taking the Fourier transform of the tangent vector, we can find an expression for the dephasing filter function~\cite{LIDesigningArbitrarySingleaxisRotations2021}:
\begin{align}
    F_z(\omega) = \frac{1}{2} \Big\|\int_0^{T_g} dt \,\vec T e^{-i\omega t}\Big\|_2^2.
\end{align}
In this section, we provide the derivation of the curve filtering index (CFI). We can study the low-frequency noise sensitivity of our control by assuming that the power spectral density has a dominant low-frequency behavior of the form
\begin{align}
    S(\omega) = \frac{\lambda^2 \,T_g}{(\omega/\omega_B)^2},
\end{align}
where $\lambda$ is the noise amplitude. In such a case, the average gate infidelity using the dephasing filter function becomes
    \begin{align}
    \mathcal{I}_n^{z} &= \frac{1}{2\pi(d+1)} \int_{-\infty}^{\infty}d\omega\, S_{z}(\omega) F_{z}(\omega) \nonumber\\
    &= \frac{\lambda^2 \omega_B^2 T_g}{2\pi(d+1)} \int_{-\infty}^{\infty}d\omega\, \frac{F_{z}(\omega)}{\omega^2}.
\end{align}
By performing an integration by parts, we can rewrite the dephasing filter function as
\begin{align}
    F_z(\omega) &= \frac{1}{2}\Big\|\int_0^{T_g}dt \, \vec T e^{-i\omega t} \Big\|_2^2 \nonumber\\
    &=\frac{1}{2} \Big\|\vec r(T_g)e^{-i\omega T_g} - \vec r(0) + i\omega \int_0^{T_g}dt \, \vec r e^{-i\omega t}  \Big\|_2^2 \nonumber\\
    &=
    \frac{\omega^2}{2} \Big\|\int_0^{T_g}dt \, \vec r e^{-i\omega t}  \Big\|_2^2,
\end{align}
where in the last step, we enforced the closed-curve condition. We are now able to exactly cancel the $\omega^2$ in the denominator, and using the definition of the delta function, we find
  \begin{align}
    \mathcal{I}_n^{z} 
    = \frac{\lambda^2 \omega_B^2 T_g}{2(d+1)} \int_{0}^{T_g}dt\, \|\vec r\|_2^2 .
\end{align}
We can now construct a geometric quantity that we call CFI by expressing the infidelity using products of dimensionless quantities:
\begin{align}
    \mathcal{I}_n^{z} 
    &= \frac{(T_g\lambda)^2 (T_g\omega_B)^2}{2(d+1)} \frac{1}{T_g^3}\int_{0}^{T_g}dt\, \|\vec r\|_2^2 \nonumber \\
    &= \frac{(T_g\lambda)^2 (T_g\omega_B)^2}{2(d+1)} \times \text{CFI},
\end{align}
where
\begin{align}
    \text{CFI} =  \frac{1}{T_g^3}\int_{0}^{T_g}dt\, \|\vec r\|_2^2.
\end{align}

For even-ordered power-law PSDs, if we find a sequence of closed curves, we can extend this quantity to higher-order exponents. This result is closely related to the findings in  Ref.~\cite{LIDesigningArbitrarySingleaxisRotations2021}, where a sequence of closed curves cancels low-frequency noise to leading order.

\section{Noise generation from a given power spectral density}
\label{noise_gen_calc}

In this section, we briefly describe the procedure to generate noise with an arbitrary PSD and specialize the equations for the production of $\omega^{-1}$ and $\omega^{-2}$ noise. The simulation of a quantum evolution is done in discrete time, therefore when we assume $N$ time points, with total evolution time $T_g$, we can define the sampling period or their distance as $T_s = \frac{T_g}{N}$. For any continuous-time signal $x(t)$, we can extract samples every $T_s$ and formally write:
\begin{align}
    x[n] = x(nT_s), n = 0,1,\dots N-1,
\end{align}
where $x[n]$ refers to the discrete-time signal. 

The first step towards the generation of a desired PSD is to connect the frequency response of the discrete-time noise signal with its continuous-time counterpart. We start by defining the Fourier transform pairs as:
\begin{align}
    x(t) \leftrightarrow X(\omega) ,\\
    x[n] \leftrightarrow X_d(\omega'),
\end{align}
where $X_d(\omega')$ is the Discrete-Time Fourier Transform (DTFT) \cite{OPPENHEIMDiscretetimeSignalProcessing1999} and $\omega'$ is the normalized (discrete-time) angular frequency. The DTFT is defined as:
\begin{align}
    X_d(\omega') = \sum_{n=-\infty}^{\infty} x[n]e^{-in\omega'},
\end{align}
The spectrum is  periodic with respect to $\omega'$, so we typically limit the analysis to the interval $|\omega'|\le \pi$. The function $X(\omega)$ is the usual continuous-time Fourier transform.

When the discrete samples are acquired from a continuous-time signal, the two frequency domains are connected by the relationship \cite{OPPENHEIMDiscretetimeSignalProcessing1999, PROAKISDigitalSignalProcessing2007,CHAPARROSignalsSystemsUsingMATLAB2019,MOUSTAKIDESVasikesTechnikesPsephiakesEpexergasias2004}
\begin{align}
    X_d(\omega') = \frac{1}{T_s} \sum_k X\left(\frac{\omega' - 2k\pi}{T_s} \right),
\end{align}
where the frequencies are related by $\omega' = \omega T_s $. When $|\omega'| \le \pi$, the DTFT contains contributions not only from the Fourier transform of the signal but also from its shifted versions. When the sampling rate is not high enough, the spectrum of the reconstructed signal is distorted, a phenomenon called \textit{aliasing}.

For aliasing-free reconstruction, we consider a band-limited signal with
\begin{align}
    X(\omega) =0, \qquad |\omega| \ge \omega_{\max}.
\end{align}
The sampling theorem states that when the sampling period is chosen such that
\begin{align}
    \frac{1}{T_s} \ge 2\times\frac{\omega_{\max}}{2\pi},
    \end{align}
the signal can be reconstructed through arbitrary interpolation with no aliasing. The minimum required sampling frequency is called \textit{Nyquist rate} \cite{OPPENHEIMDiscretetimeSignalProcessing1999, PROAKISDigitalSignalProcessing2007,CHAPARROSignalsSystemsUsingMATLAB2019,MOUSTAKIDESVasikesTechnikesPsephiakesEpexergasias2004}.
In a computer simulation, every ODE solver interpolates the signal samples in a distinct manner. The simplest method is to maintain the sample value for time $T_s$ (zero-order hold) \cite{CHAPARROSignalsSystemsUsingMATLAB2019}. Assuming that the signal is ideally reconstructed when $T_s$ is adequately small, in accordance to the Nyquist-Shannon criterion, we can still maintain that
\begin{align}
    X_s(\omega) = T_s \,X_d(\omega T_s),\qquad |\omega| \le \frac{\pi}{T_s} = N\frac{\pi}{T_g},
    \label{sim_spectra}
\end{align}
where $X_s$ is the Fourier Transform of the signal in the simulation.

Now that we have established the continuous-discrete connection, we turn our attention to the noise generation procedure. Let us assume that we are interested in adding white noise in our simulation. We therefore need to produce a constant $S(\omega)$. Drawing random samples that correspond to an autocorrelation function of the form
\begin{align}
    R_d[n-m] = \sigma_d^2 \delta_{nm},
\end{align}
will in turn produce a PSD of the form $S_d(\omega') = \sigma_d^2$, where $\sigma_d^2$ is the variance (over noise realizations) of the samples. In order to construct more elaborate PSDs, we study such a procedure in the context of  signal filtering \cite{OPPENHEIMDiscretetimeSignalProcessing1999, PROAKISDigitalSignalProcessing2007,CHAPARROSignalsSystemsUsingMATLAB2019, MOUSTAKIDESVasikesTechnikesPsephiakesEpexergasias2004}. In systems theory, driving a system using $x[n]$, will yield an output $y[n]$ that can be computed in the (normalized) frequency domain as
\begin{align}
    Y_d(\omega') = H_d(\omega') X_d(\omega'),
\end{align}
where $H_d(\omega')$ is the discrete-time system transfer function. It can be shown \cite{KASDINDiscreteSimulationColoredNoise1995, PAPOULISProbabilityRandomVariablesStochastic2002} that  the PSDs of the input and output processes are related by
\begin{align}
    S_{yd}(\omega') = |H_d(\omega')|^2 S_{xd}(\omega').
\end{align}
Consequently, if we drive the system with white noise, the transfer function can be constructed in such a way that the output samples follow the desired PSD. The filtering process can be carried out in the time domain, where the calculation of the output is found as the convolution of the input with the discrete impulse response. Formally, we can compute
\begin{align}
    h[n] = \frac{1}{2\pi} \int_{-\pi}^{\pi} d\omega'H_d(\omega') e^{in\omega'},
\end{align}
but in practice, we can derive a difference equation that provides the output samples recursively. Other implementations involve using the Fast Fourier Transform (FFT) to calculate the output of the system.

For this work, we are interested in generating $\omega^{-\alpha}$ noise. We will outline the procedure described in Ref.~\cite{KASDINDiscreteSimulationColoredNoise1995}. The generation of the discrete PSD  must simulate a function of the form
\begin{align}
    S_d(\omega') = \frac{\sigma_d^2 }{(\omega')^\alpha}.
\end{align}
Assuming a general form of the noise PSD:
\begin{align}
    S(\omega) = \frac{\lambda^2 T_g}{(\omega/\omega_B)^\alpha}, \qquad \omega_B = \frac{2\pi}{T_g},
\end{align}
we will set the input noise variance according to the relationship of the spectra from Eq.~\eqref{sim_spectra} as
\begin{align}
    \sigma_d^2 = (\lambda\sqrt{N})^2 (\frac{2\pi}{N})^\alpha .
\end{align}
A valid approximation can be achieved with a transfer function of the form \cite{KASDINDiscreteSimulationColoredNoise1995}:
\begin{align}
  H_d(\omega') = \frac{1}{(1-e^{i\omega'})^\frac{\alpha}{2}} .
\end{align}
In order for this transfer function to be realized, a Finite Impulse Response (FIR) filter is employed with impulse response
\begin{align}
    h[n] = \frac{\Gamma(\frac{\alpha}{2} + n)}{n!\Gamma(\alpha/2)}.
\end{align}
The implementation of this method can be found in our repository, in the file \texttt{noisetools.py}. 
A general treatment of the signal filtering techniques can be found in Ref.~\cite{OPPENHEIMDiscretetimeSignalProcessing1999, PROAKISDigitalSignalProcessing2007,CHAPARROSignalsSystemsUsingMATLAB2019, MOUSTAKIDESVasikesTechnikesPsephiakesEpexergasias2004}.

When time-dependent noise is generated, there are two important elements that need to be addressed to ensure that the predictions involving the filter functions are correct. The first issue arises in the filtering process and concerns the response of the FIR filter. Depending on the order of the FIR, the moments of the output signal become stationary only after a few initial samples, as a result of the filter's transient effects. The second issue is about the validation of the noise PSD. There are several methods that estimate the PSD from noise samples \cite{KAYModernSpectralEstimationTheory1988} and some of them involve the usage of the Fourier transform. Since the FFT is applied on a truncated sequence of the otherwise infinite-sample signal, a windowing process takes place that distorts the output spectrum. In that case, low-frequency PSDs might appear with larger amplitudes when a rectangular window is used for spectrum estimation \cite{KASDINDiscreteSimulationColoredNoise1995}. When such an issue arises, a simple solution is to estimate the derivative of the output signal. From the properties of the Fourier Transform, the PSD will contain a factor of $\omega^2$, capable of eliminating the low-frequency behavior that creates such a distortion. Obtaining a better estimate of the PSD by filtering the input samples is often encountered in the literature as \textit{prewhitening} \cite{KAYModernSpectralEstimationTheory1988}.

%\nocite{*}

\bibliography{Qurveros}% Produces the bibliography via BibTeX.

\end{document}